\documentclass[pdftex,smallextended,epjc3]{svjour3}
\RequirePackage[T1]{fontenc}

\smartqed

\RequirePackage{graphicx}
\RequirePackage{mathptmx}
\RequirePackage{flushend}
\RequirePackage[numbers,sort&compress]{natbib}
\RequirePackage[colorlinks,citecolor=blue,urlcolor=blue,linkcolor=blue]{hyperref}

\usepackage{amsmath,amssymb}

\newcommand{\ind}[2]{^{#1}_{\text{#2}}}
\newcommand{\inds}[2]{^{#1}_{\text{\scriptsize{#2}}}}
\newcommand{\indt}[2]{^{#1}_{\text{\tiny{#2}}}}
\newcommand{\ARe}[1]{a\inds{#1}{Re}}
\newcommand{\AIm}[1]{a\inds{#1}{Im}}
\newcommand{\ATL}[2]{A^{#1}_{\text{\scriptsize TL},#2}}
\newcommand{\ATLs}[2]{A^{#1}_{\text{\tiny TL},#2}}
\def\mpi{m_{\pi}}
\def\DP{\Delta\Pi}
\def\nf{n_{f}}
\def\co{\mathcal{O}}
\def\makeheadbox{}

\begin{document}

\title{Electron--positron annihilation into hadrons at the higher--loop levels}

\author{A.V.~Nesterenko\thanksref{e1,addr1}}

\thankstext{e1}{e-mail: nesterav@theor.jinr.ru}

\institute{Bogoliubov Laboratory of Theoretical Physics,
Joint Institute for Nuclear Research,
Dubna, 141980, Russian Federation\label{addr1}}

\date{~}

\maketitle

\begin{abstract}
The strong corrections to the $R$--ratio of electron--positron
annihilation into hadrons are studied at the higher--loop levels.
Specifically, the essentials of continuation of the spacelike perturbative
results into the timelike domain are elucidated. The derivation of a
general form of the commonly employed approximate expression for the
$R$--ratio (which constitutes its truncated re--expansion at high
energies) is delineated, the appearance of the pertinent $\pi^2$--terms is
expounded, and their basic features are examined. It~is demonstrated that
the validity range of such approximation is strictly limited to
$\sqrt{s}/\Lambda > \exp(\pi/2) \simeq 4.81$ and that it converges rather
slowly when the energy scale approaches this value. The~spectral function
required for the proper calculation of the $R$--ratio is explicitly
derived and its properties at the higher--loop levels are studied.
The~developed method of calculation of the spectral function enables one
to obtain the explicit expression for the latter at an arbitrary loop
level. By making use of the derived spectral function the proper
expression for the $R$--ratio is calculated up to the five--loop level and
its properties are examined. It~is shown that the loop convergence of the
proper expression for the $R$--ratio is better than that of its commonly
employed approximation. The impact of the omitted higher--order
$\pi^2$--terms on the latter is also discussed.
\end{abstract}

\section{Introduction}
\label{Sect:Intro}

In the studies of a variety of the strong interaction processes a key role
is played by the hadronic vacuum polarization function~$\Pi(q^2)$, the
related function~$R(s)$, and the Adler function~$D(Q^2)$. In~particular,
these functions govern such processes as the electron--positron
annihilation into hadrons, inclusive hadronic decays of $\tau$~lepton and
$Z$~boson, as well as the hadronic contributions to precise electroweak
observables, such as the muon anomalous magnetic moment
\mbox{$(g-2)_{\mu}$} and the running of the electromagnetic fine structure
constant. The~theoretical analysis of these processes constitutes a
decisive self--consistency test of Quantum Chromodynamics~(QCD) and entire
Standard Model, that, in turn, puts robust restrictions on a possible New
Physics beyond the latter. Additionally, the energy scales relevant to the
foregoing strong interaction processes span from the infrared to
ultraviolet domain, so that their theoretical investigation provides a
native framework for a profound study of both perturbative and
intrinsically nonperturbative aspects of hadron dynamics. It~is worth
noting also that a majority of the aforementioned processes are of a
direct relevance to the physics at the currently designed Future Collider
Projects, such as the Future Circular Collider FCC--ee~\cite{FCCee},
Circular Electron--Positron Collider CEPC (its first phase)~\cite{CEPC},
the International Linear Collider ILC~\cite{ILC}, the Compact Linear
Collider CLIC~\cite{CLIC}, as well as the E989~experiment at
Fermilab~\cite{E989}, the E34~experiment at~\mbox{J--PARC}~\cite{E34}, and
others.

In fact, over the past decades the perturbative approach to QCD remains a
basic tool for the theoretical exploration of the hadronic physics.
However, the QCD~perturbation theory can be directly applied to the study
of the strong interaction processes only in the spacelike (Euclidean)
domain, whereas the proper description of hadron dynamics in the timelike
(Minkowskian) domain additionally requires the pertinent dispersion
relations. Specifically, the dispersion relation for the $R$--ratio of
electron--positron annihilation into hadrons converts the physical
kinematic restrictions on the process on hand into the mathematical form
and determines the way how the ``timelike'' observable~$R(s)$ is related
to the ``spacelike'' quantity~$D(Q^2)$, the corresponding perturbative
input being embodied by the so--called spectral function. Since the
calculation of the latter at the higher--loop levels constitutes a rather
challenging task, one commonly resorts to an approximate form of the
$R$--ratio, namely, its truncated re--expansion at high energies. At the
same time, one has to be aware that at any given loop level such
re--expansion generates an infinite number of the so--called
$\pi^2$--terms (which may not necessarily be small enough to be safely
discarded at the higher orders), that also worsen the loop convergence of
the resulting approximate~$R$--ratio.

The primary objective of the paper is to explicitly derive a general form
of the spectral function required for the proper evaluation of the
$R$--ratio and to study its properties up to the five--loop level. It is
also of an apparent interest to calculate the $R$--ratio itself, to
examine its higher--loop convergence, and to elucidate the impact of the
omitted higher--order $\pi^2$--terms on its truncated re--expanded
approximation.

The layout of the paper is as follows. In~Sect.~\ref{Sect:Repem} the
essentials of continuation of the spacelike perturbative results into the
timelike domain are expounded. In~Sect.~\ref{Sect:Repem1L} the one--loop
expression for the $R$--ratio is explicated and its approximations are
discussed. In~Sect.~\ref{Sect:RepemPi2} the derivation of a general form
of the commonly employed approximate expression for the $R$--ratio (which
constitutes its truncated re--expansion at high energies) is delineated,
the appearance of the pertinent $\pi^2$--terms is elucidated, and their
basic features are studied. In~Sect.~\ref{Sect:RhoHL} the explicit form of
the spectral function required for the proper calculation of the
$R$--ratio is obtained and its properties at the higher--loop levels are
examined. By making use of the derived spectral function in
Sect.~\ref{Sect:RepemHL} the proper expression for the $R$--ratio is
calculated up to the five--loop level and its properties are studied.
Additionally, the obtained $R$--ratio is juxtaposed with its commonly
employed approximation and the impact of the omitted higher--order
$\pi^2$--terms on the latter is discussed. In~the Conclusions
(Sect.~\ref{Sect:Concl}) the basic results are summarized.

\section{$R$--ratio of electron--positron annihilation into hadrons}
\label{Sect:Repem}

As noted above, the theoretical analysis of certain strong interaction
processes relies on the hadronic vacuum polarization function~$\Pi(q^2)$,
which is defined as the scalar part of the hadronic vacuum polarization
tensor
\begin{eqnarray}
\label{P_Def}
\Pi_{\mu\nu}(q^2) &=& i\!\int\!d^4x\,e^{i q x} \bigl\langle 0 \bigl|\,
T\!\left\{J_{\mu}(x)\, J_{\nu}(0)\right\} \bigr| 0 \bigr\rangle =
\nonumber \\
&=& \frac{i}{12\pi^2} (q_{\mu}q_{\nu} - g_{\mu\nu}q^2) \Pi(q^2).
\end{eqnarray}
For the processes involving final state hadrons the
function~$\Pi(q^2)$~(\ref{P_Def}) has the only cut along the positive
semiaxis of real~$q^2$ starting at the hadronic production threshold~$q^2
\ge 4\mpi^2$ (the discussion of this issue can be found in, e.g.,
Ref.~\cite{Feynman}). In particular, the Feynman amplitude of the
respective process vanishes for the energies below the threshold, that
expresses the physical fact that the production of the final state hadrons
is kinematically forbidden for~$q^2 < 4\mpi^2$. In turn, the known
location of the cut of function~$\Pi(q^2)$ in the complex $q^2$~plane
enables one to write down the pertinent dispersion relation
\begin{equation}
\label{P_Disp}
\DP(q^2\!,\, q_0^2) = (q^2 - q_0^2) \int\limits_{4\mpi^2}^{\infty}
\frac{R(\sigma)}{(\sigma-q^2)(\sigma-q_0^2)}\, d\sigma,
\end{equation}
with the once--subtracted Cauchy integral formula being employed.
In~Eq.~(\ref{P_Disp}) $\DP(q^2\!,\, q_0^2) = \Pi(q^2) - \Pi(q_0^2)$,
whereas $R(s)$~stands for the discontinuity of the hadronic vacuum
polarization function across the physical cut
\begin{equation}
\label{R_Def}
R(s) = \frac{1}{2 \pi i} \lim_{\varepsilon \to 0_{+}}
\DP(s+i\varepsilon,s-i\varepsilon).
\end{equation}
This function is commonly identified with the so--called $R$--ratio of
electron--positron annihilation into hadrons $R(s) = \sigma(e^{+}e^{-}
\!\to \text{hadrons}; s)/\sigma(e^{+}e^{-} \!\to \mu^{+}\mu^{-}; s)$, with
$s=q^2>0$ being the timelike kinematic variable, namely, the
center--of--mass energy squared.

\begin{figure}[t]
\centering
\includegraphics[width=75mm,clip]{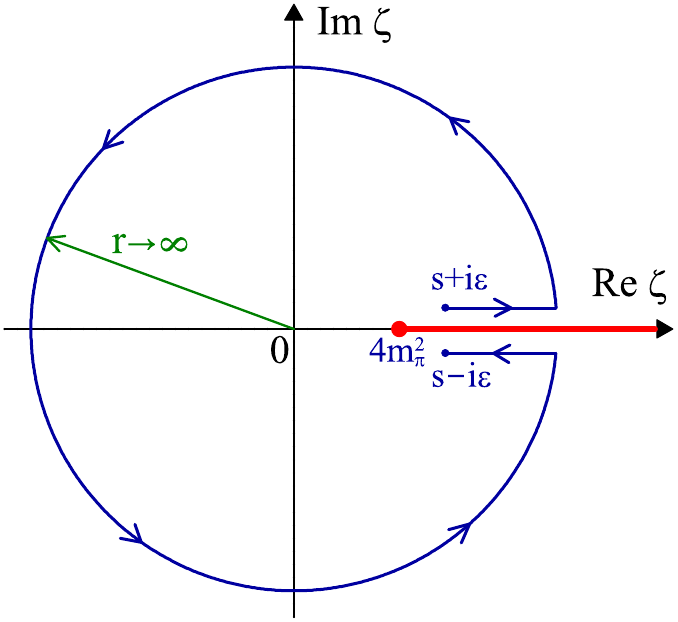}
\caption{The integration contour in Eq.~(\ref{R_Disp2}). The physical cut
$\zeta \ge 4\mpi^2$ of the Adler function $D(-\zeta)$~(\ref{Adler_Def}) is
shown along the positive semiaxis of real~$\zeta$.}
\label{Plot:Contour}
\end{figure}

In practice one deals with the Adler function~$D(Q^2)$~\cite{Adler}, which
is defined as the logarithmic derivative of the hadronic vacuum
polarization function~(\ref{P_Def})
\begin{equation}
\label{Adler_Def}
D(Q^2) = - \frac{d\, \Pi(-Q^2)}{d \ln Q^2},
\end{equation}
with $Q^2 = -q^2 > 0$ being the spacelike kinematic variable. Note that
the subtraction point~$q_0^2$ entering Eq.~(\ref{P_Disp}) does not appear
in Eqs.~(\ref{R_Def}) and~(\ref{Adler_Def}). The widely employed
dispersion relation for the Adler function follows immediately from
Eqs.~(\ref{P_Disp}) and~(\ref{Adler_Def}), specifically~\cite{Adler}
\begin{equation}
\label{Adler_Disp}
D(Q^2) = Q^2 \int\limits_{4\mpi^2}^{\infty}
\frac{R(\sigma)}{(\sigma+Q^2)^2}\, d\sigma.
\end{equation}
In particular, this dispersion relation enables one to extract the
experimental prediction for the Adler function by making use of the
corresponding experimental data on electron--positron annihilation into
hadrons. However, to obtain the theoretical expression for the~$R$--ratio
itself, the relation inverse to Eq.~(\ref{Adler_Disp}) is required. The
latter can be obtained by integrating Eq.~(\ref{Adler_Def}) in finite
limits, that yields~\cite{Rad82, KP82}
\begin{equation}
\label{R_Disp2}
R(s) =  \frac{1}{2 \pi i} \lim_{\varepsilon \to 0_{+}}
\int\limits_{s + i \varepsilon}^{s - i \varepsilon}
D(-\zeta)\,\frac{d \zeta}{\zeta}.
\end{equation}
Specifically, this equation relates the $R$--ratio to the theoretically
calculable Adler function and provides a native way to properly account
for the effects due to continuation of the spacelike perturbative results
into the timelike domain. The integration contour in Eq.~(\ref{R_Disp2})
lies in the region of analyticity of the integrand, see
Fig.~\ref{Plot:Contour}. Note also that the relation, which expresses the
hadronic vacuum polarization function in terms of the Adler function can
be obtained in a similar way, namely~\cite{Pivovarov91}
\begin{equation}
\label{P_Disp2}
\DP(-Q^2\!,\, -Q_0^2) = - \int\limits_{Q_0^2}^{Q^2} D(\zeta)
\frac{d \zeta}{\zeta},
\end{equation}
where~$Q^{2}$ and~$Q_{0}^{2}$ stand for the spacelike kinematic variable
and the subtraction point, respectively. Basically,
Eqs.~(\ref{P_Disp})--(\ref{P_Disp2}) constitute the complete set of
relations, which express the functions on hand in terms of each other.
It~is worth mentioning here that, in general, the pattern of applications
of the dispersion relations in theoretical particle physics is quite
diverse. For example, among the latter are such issues as the refinement
of chiral perturbation theory~\cite{DRChPT, Passemar}, the accurate
determination of parameters of resonances~\cite{DRRes}, the assessment of
the hadronic light--by--light scattering~\cite{Colangelo}, as well as many
others.

It is necessary to outline that the derivation of dispersion
relations~(\ref{P_Disp})--(\ref{P_Disp2}) is based only on the kinematics
of the process on hand and involves neither model--dependent
phenomenological assumptions nor additional approximations. In~turn, the
relations~(\ref{P_Disp})--(\ref{P_Disp2}) impose a number of strict
physical inherently nonperturbative constraints on the
functions~$\Pi(q^2)$, $R(s)$, and~$D(Q^2)$, that should definitely be
taken into account when one intends to go beyond the limits of
applicability of the QCD perturbation theory. It is worthwhile to note
that these nonperturbative restrictions have been merged with the
corresponding perturbative input in the framework of dispersively improved
perturbation theory~(DPT)~\cite{Book, DPT1, DPT2} (its preliminary
formulation was discussed in Ref.~\cite{DPTPrelim}). In particular, the
DPT enables one to overcome some intrinsic difficulties of the QCD
perturbation theory and to extend its applicability range towards the
infrared domain, see Ref.~\cite{Book} and references therein for the
details.

In the framework of perturbation theory the Adler
function~(\ref{Adler_Def}) takes the form of the power series in the
so--called QCD couplant $a\ind{(\ell)}{s}(Q^2) =
\alpha\ind{(\ell)}{s}(Q^2)\, \beta_{0}/(4\pi)$, namely
\begin{equation}
\label{AdlerPert}
D^{(\ell)}_{{\rm pert}}(Q^2) = 1 + d^{(\ell)}_{{\rm pert}}(Q^2),
\quad
d^{(\ell)}_{{\rm pert}}(Q^2) = \!\!\sum_{j=1}^{\ell}\! d_{j}\!
\left[a\ind{(\ell)}{s}(Q^2)\!\right]^{j}\!.
\end{equation}
In this equation~$\ell$ specifies the loop level, $d_1 = 4/\beta_0$,
$\beta_0 = 11 - 2\nf/3$, $\nf$~is the number of active flavors, and
the~common prefactor $N_{\text{c}}\sum_{f=1}^{\nf} Q_{f}^{2}$ is omitted
throughout, where $N_{\text{c}}=3$ denotes the number of colors and
$Q_{f}$~stands for the electric charge of $f$--th quark. The QCD couplant
$a\ind{(\ell)}{s}(Q^2)$ entering Eq.~(\ref{AdlerPert}) can be represented
as the double sum
\begin{equation}
\label{AItGen}
a\ind{(\ell)}{s}(Q^2) =
\sum_{n=1}^{\ell}\sum_{m=0}^{n-1} b^{m}_{n}\,
\frac{\ln^{m}(\ln z)}{\ln^n z},
\end{equation}
where $z=Q^2/\Lambda^2$ and~$b^{m}_{n}$ (the integer superscript~$m$ is
not to be confused with respective power) stands for the combination of
the $\beta$~function perturbative expansion coefficients, specifically,
$b^{0}_{1}=1$, $b^{0}_{2}=0$, $b^{1}_{2}=-\beta_{1}/\beta^{2}_{0}$,~etc.
The Adler function perturbative expansion coefficients~$d_{j}$ were
calculated up to the four--loop level ($1 \le j \le 4$), see
Ref.~\cite{RPert4L} and references therein, whereas for the five--loop
coefficient~$d_{5}$ only numerical estimation~\cite{RPert5LEstim1} is
available so~far. The numerical values of the perturbative
coefficients~$d_{j}$~(\ref{AdlerPert}) are listed in
Tab.~\ref{Tab:AdlerPert}. In turn, the $\beta$~function perturbative
expansion coefficients~$\beta_{j}$ have been calculated up to the
five--loop level ($0 \le j \le 4$), see Ref.~\cite{Beta5L} and references
therein for the details.

\begin{table*}[t]
\caption{Numerical values of the Adler function perturbative expansion
coefficients~$d_j$~(\ref{AdlerPert}). In~the last column the numerical
estimation of the five--loop coefficient~$d_5$~\cite{RPert5LEstim1} is
listed.}
\label{Tab:AdlerPert}
\begin{tabular*}{\textwidth}{@{\extracolsep{\fill}}cccccc@{\extracolsep{\fill}}}
\hline
$\nf$ & $d_{1}$ & $d_{2}$ & $d_{3}$ & $d_{4}$ & $d_{5}$ \\
\hline
0 & 0.3636 & 0.2626 &   0.8772  &   2.3743  & \textit{5.40}   \\
1 & 0.3871 & 0.2803 &   0.7946  &   2.1884  & \textit{4.70}   \\
2 & 0.4138 & 0.3005 &   0.7137  &   2.1466  & \textit{3.74}   \\
3 & 0.4444 & 0.3239 &   0.5593  &   1.9149  & \textit{2.52}   \\
4 & 0.4800 & 0.3513 &   0.2868  &   1.3440  & \textit{1.16}   \\
5 & 0.5217 & 0.3836 & $-0.1021$ &   0.6489  & \textit{0.0256} \\
6 & 0.5714 & 0.4225 & $-0.7831$ & $-0.8952$ & \textit{0.267}  \\
\hline
\end{tabular*}
\end{table*}

In what follows the nonperturbative aspects of the strong interactions
will be disregarded and a primary attention will be given to the
theoretical description of the $R$--ratio of electron--positron
annihilation into hadrons at moderate and high energies.  For this purpose
the effects due to the masses of the involved particles can be safely
neglected (a discussion of the impact of such effects\footnote{For
example, in the limit~$\mpi=0$ some of the aforementioned nonperturbative
constraints on the functions on hand appear to be lost.} on the
low--energy behavior of the functions~$\Pi(q^2)$, $R(s)$, and~$D(Q^2)$ can
be found in, e.g., Refs.~\cite{Book, DPT1, DPT2, QCD8245, NPQCD71, C246}).
Additionally, for the scheme--dependent perturbative
coefficients~$\beta_{j}$ and~$d_{j}$ the $\overline{\rm{MS}}$--scheme will
be assumed and for the uncalculated yet five--loop coefficient~$d_5$ its
numerical estimation~\cite{RPert5LEstim1} will be employed.

Thus, in the massless limit the relation~(\ref{R_Disp2}) can be
represented as (see also Ref.~\cite{APT0a})
\begin{equation}
\label{RProp}
R^{(\ell)}(s) = 1 + r^{(\ell)}(s),
\quad
r^{(\ell)}(s) =
\int\limits_{s}^{\infty}\!\rho^{(\ell)}(\sigma)\,
\frac{d \sigma}{\sigma},
\end{equation}
where
\begin{equation}
\label{RhoDef}
\rho^{(\ell)}(\sigma) =
\frac{1}{2 \pi i} \lim_{\varepsilon \to 0_{+}}
\Bigl[d^{(\ell)}(-\sigma - i \varepsilon) -
d^{(\ell)}(-\sigma + i \varepsilon)\!\Bigr]
\end{equation}
stands for the spectral function and~$d^{(\ell)}(Q^2)$ denotes the
$\ell$--loop strong correction to the Adler function. As~mentioned above,
only perturbative contributions will be retained in Eq.~(\ref{RhoDef})
hereinafter, that makes Eq.~(\ref{RProp}) identical to that of both the
foregoing DPT~\cite{Book, DPT1, DPT2} and the so--called analytic
perturbation theory\footnote{The discussion of APT and its applications
can be found in, e.g., Refs.~\cite{APT0a, APT0b, APT1, APT2, APT3, APT4,
APT5, APT6, APT7a, APT7b, APT8a, APT8b, APT9, APT10}.}~(APT)~\cite{APT0a,
APT0b}. It~has to be noted that, in general, the perturbative spectral
function at small values of its argument may be altered by the terms of an
intrinsically nonperturbative nature. For instance, the nonperturbative
models discussed in Refs.~\cite{PRD6264, Review, MPLA1516, APTCSB}
constitute a superposition of the perturbative spectral function with the
so--called ``flat''~terms (which by definition do not affect the
corresponding perturbative results at high energies), whereas the
models~\cite{12dAnQCD} modify the low--energy behavior of the perturbative
spectral function proceeding from certain phenomenological assumptions.

It is worthwhile to mention also that a ``naive'' approach to continue the
spacelike perturbative result~(\ref{AdlerPert}) into the timelike domain
consists in merely identifying the timelike kinematic variable~($s=q^2$)
with the spacelike one~($Q^2=-q^2$),~i.e.,
\begin{equation}
\label{RNaive}
R\ind{(\ell)}{naive}(s) = D\ind{(\ell)}{pert}(|s|) =
1 + \!\sum_{j=1}^{\ell}\! d_{j}\!
\left[a\ind{(\ell)}{s}(|s|)\!\right]^{j}\!.
\end{equation}
However, as thoroughly discussed in Ref.~\cite{Penn}, this prescription
yields a misleading result, which differs from the proper
one~(\ref{RProp}) even in the deep ultraviolet asymptotic, see also
Refs.~\cite{Pi2Terms1, ProsperiAlpha, RPert5LEstim1} as well
as~\cite{Book} and references therein.

\section{$R$--ratio at the one--loop level}
\label{Sect:Repem1L}

\begin{figure}[t]
\centering
\includegraphics[width=75mm,clip]{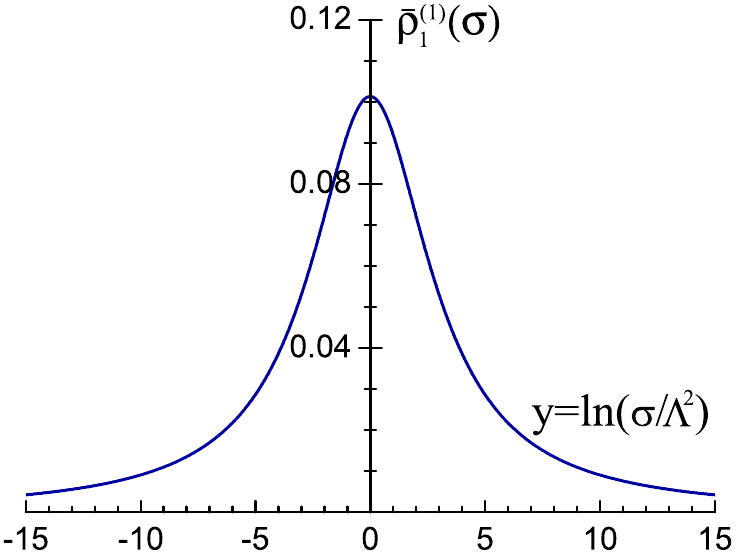}
\caption{The one--loop spectral
function~$\bar\rho^{(1)}_{1}(\sigma)$~(\ref{RhoPert1L}).}
\label{Plot:RhoY1L}
\end{figure}

Let us address now the $R$--ratio of electron--positron annihilation into
hadrons at the one--loop level. As discussed in the previous Section, for
this purpose the spectral function~$\rho(\sigma)$~(\ref{RhoDef}), which
enters the pertinent integral representation~(\ref{RProp}), is required.
Since the involved strong correction to the Adler
function~(\ref{AdlerPert}) takes a simple form at the one--loop level
\begin{equation}
\label{DscPert1L}
d\ind{(1)}{pert}(Q^2) = d_{1}\,a\ind{(1)}{s}(Q^2),
\quad
a\ind{(1)}{s}(Q^2) = \frac{1}{\ln(Q^2/\Lambda^2)}
\end{equation}
and
\begin{equation}
\label{LnCut}
\lim_{\varepsilon \to 0_{+}} \ln(x \pm i\varepsilon) = \ln |x| \pm i\pi\theta(-x),
\end{equation}
the calculation of~$\rho^{(1)}(\sigma)$~(\ref{RhoDef}) appears to be
quite straightforward. Specifically, the one--loop spectral
function~(\ref{RhoDef}) for the positive values of its argument reads
\begin{equation}
\rho^{(1)}(\sigma) = d_{1}\,\frac{1}{2 \pi i}
\left[\frac{1}{\ln(\sigma/\Lambda^2) - i \pi} -
\frac{1}{\ln(\sigma/\Lambda^2) + i \pi}\right]\!\!,
\end{equation}
that, in turn, can be represented as
\begin{equation}
\label{RhoPert1L}
\rho^{(1)}(\sigma) = d_{1} \bar\rho^{(1)}_{1}(\sigma),
\quad
\bar\rho^{(1)}_{1}(\sigma) = \frac{1}{y^2+\pi^2},
\quad
y=\ln\!\biggl(\!\frac{\sigma}{\Lambda^2}\!\biggr)\!.
\end{equation}
In these equations~$\theta(x)$ is the Heaviside unit step function
[i.e.,~$\theta(x)=1$ if  $x \ge 0$ and~$\theta(x)=0$ otherwise],
$d_{1}=4/\beta_{0}$, and~$\beta_{0}=11 - 2\nf/3$. The~plot of the
one--loop spectral function~$\bar\rho^{(1)}_{1}(\sigma)$~(\ref{RhoPert1L})
is displayed in Fig.~\ref{Plot:RhoY1L}. As one can infer from this Figure,
the function on hand assumes the values in the interval $0 \le
\bar\rho^{(1)}_{1}(\sigma) \le 1/\pi^2$ and decreases as~$1/y^2$ in both
ultraviolet~($y\to\infty$) and infrared~($y\to-\infty$) asymptotics.

Then, the corresponding one--loop strong correction to the $R$--ratio can
also be easily obtained in an explicit form. Specifically, the
integration~(\ref{RProp}) of the one--loop spectral
function~(\ref{RhoPert1L}) yields\footnote{It is assumed that~$\arctan(x)$
is a monotone nondecreasing function of its argument: $-\pi/2 \le
\arctan(x) \le \pi/2$ for $-\infty < x < \infty$.}
\begin{equation}
\label{RCTL1L}
r^{(1)}(s) = d_{1}a\inds{(1)}{TL}(s),
\quad
a\inds{(1)}{TL}(s)=
\frac{1}{2} -
\frac{1}{\pi}\arctan\!\left(\frac{\ln w}{\pi}\right)\!,
\end{equation}
where~$w=s/\Lambda^2$. The function~$a\inds{(1)}{TL}(s)$~(\ref{RCTL1L})
constitutes the one--loop couplant, which properly accounts for the
effects due to continuation of the spacelike perturbative
expression~(\ref{DscPert1L}) into the timelike domain. It~is worthwhile to
note here that Eq.~(\ref{RCTL1L}) has first appeared in
Ref.~\cite{Schrempp80} and only afterwards was derived in
Refs.~\cite{Rad82, Pivovarov91, APT0a}.

\begin{figure}[t]
\centering
\raisebox{3.4mm}{\includegraphics[width=72.5mm,clip]{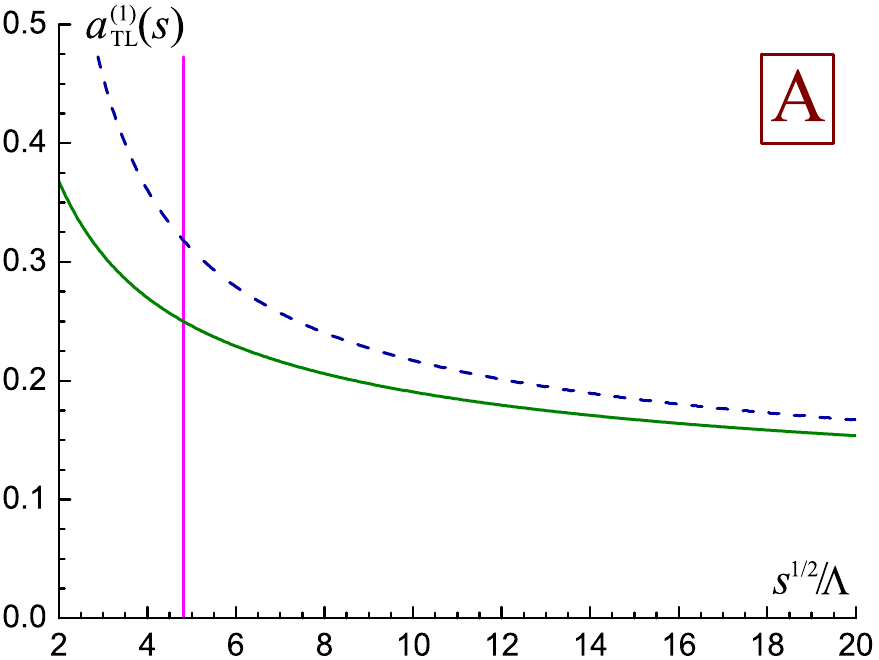}}%
\hspace{7.5mm}%
\includegraphics[width=72.5mm,clip]{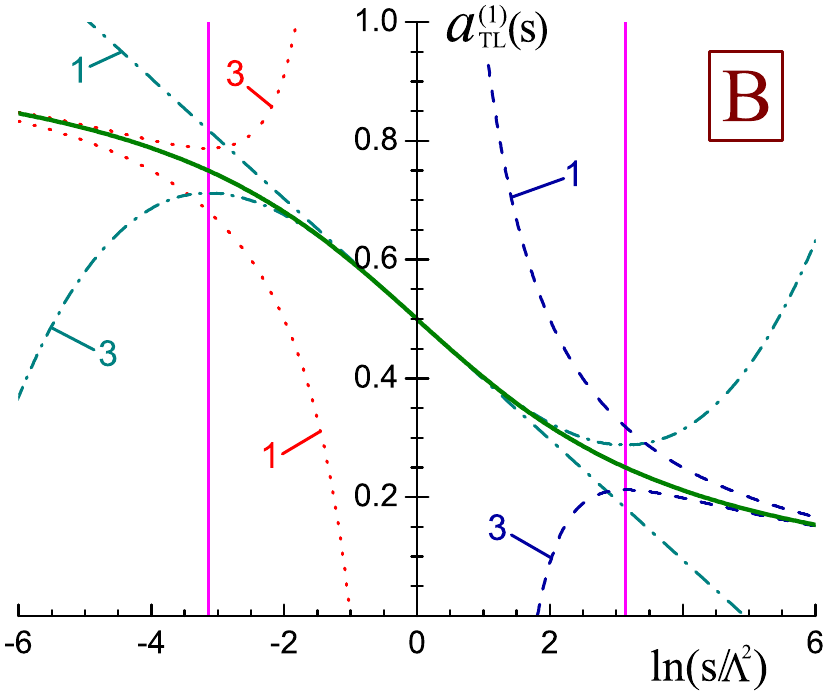}
\caption{Plot~A: The one--loop ``timelike'' effective
couplant~$a\indt{(1)}{TL}(s)$ [Eq.~(\ref{RCTL1L}), solid curve] and the
naive continuation of the one--loop perturbative couplant into the
timelike domain~$a\ind{(1)}{s}(|s|)$ [Eq.~(\ref{DscPert1L}), dashed
curve]. Plot~B: The one--loop ``timelike'' effective
couplant~$a\indt{(1)}{TL}(s)$ [Eq.~(\ref{RCTL1L}), solid curve] and its
re--expansions at various energy intervals: $\ln w < -\pi$
[Eq.~(\ref{ATL1L1pIR}), dotted curves], $-\pi < \ln w < \pi$
[Eq.~(\ref{ATL1L1pMD}), dot--dashed curves], and~$\ln w > \pi$
[Eq.~(\ref{ATL1L1pUV}), dashed curves]. Numerical labels indicate the
highest absolute value of the power of~$\ln w$ retained in the
re--expan\-si\-ons on hand. The boundaries of the convergence ranges of
Eqs.~(\ref{ATL1L1pIR})--(\ref{ATL1L1pUV}) are marked by vertical solid
lines.}
\label{Plot:ATL1L1p}
\end{figure}

Figure~\ref{Plot:ATL1L1p}$\,$A displays the one--loop ``timelike''
effective couplant~$a\inds{(1)}{TL}(s)$~(\ref{RCTL1L}) and the ``naive''
continuation of the one--loop perturbative
couplant~$a\ind{(1)}{s}(Q^2)$~(\ref{DscPert1L}) into the timelike
domain~(\ref{RNaive}). As one can infer from this Figure, at high energies
the two couplants approach each other. At the same time, at moderate
energies the deviation between the functions on hand becomes significant,
whereas in the infrared domain their behavior turns out to be
qualitatively different. Specifically, the
function~$a\ind{(1)}{s}(|s|)$~(\ref{DscPert1L}) diverges at low energies
due to the infrared unphysical singularities,
whereas~$a\inds{(1)}{TL}(s)$~(\ref{RCTL1L}) is a smooth monotone
decreasing function of its argument, which contains no singularities
for~$s>0$.

A somewhat simpler but approximate form of the strong correction to the
$R$--ratio can be obtained by its further re--expansion. Specifically, for
this purpose one splits the entire energy range~$0 < s < \infty$ into
three intervals (namely, $\ln w < -\pi$, $-\pi < \ln w < \pi$, and~$\ln w
> \pi$) and applies the Taylor expansion to~$r(s)$ in each of those
intervals. At~the one--loop level the implementation of these steps for
the expression~(\ref{RCTL1L}) yields
\begin{align}
\label{ATL1L1pIR}
a\inds{(1)}{TL}(s) & \simeq 1 + \frac{1}{\ln w} - \frac{1}{3}\frac{\pi^2}{\ln^{3}w}
+\co\!\!\left(\!\frac{1}{\ln^{5}w}\!\right)\!\!, \quad \ln w < -\pi, \\
\label{ATL1L1pMD}
a\inds{(1)}{TL}(s) & \simeq \frac{1}{2} - \frac{\ln w}{\pi^2} +
\frac{1}{3}\frac{\ln^{3}w}{\pi^4}
+\co\!\Bigl(\!\ln^{5}w\!\Bigr)\!,
\quad -\pi < \ln w < \pi, \\
\label{ATL1L1pUV}
a\inds{(1)}{TL}(s) & \simeq \frac{1}{\ln w} - \frac{1}{3}\frac{\pi^2}{\ln^{3}w}
+\co\!\!\left(\!\frac{1}{\ln^{5}w}\!\right)\!\!,
\quad \ln w > \pi,
\end{align}
where~$w=s/\Lambda^2$. In particular, as one can infer from
Fig.~\ref{Plot:ATL1L1p}$\,$B, the
re--expansions~(\ref{ATL1L1pIR})--(\ref{ATL1L1pUV}) may provide an
accurate approximation of the function~(\ref{RCTL1L}) in the
aforementioned energy intervals, if the number of retained terms is large
enough. However, as one can note, the convergence of the
re--expan\-si\-ons~(\ref{ATL1L1pIR})--(\ref{ATL1L1pUV}) becomes worse when
the energy scale approaches the delimiting values~$\sqrt{s}/\Lambda =
\exp(\pm\pi/2)$, see also discussion of this issue in
Sect.~\ref{Sect:RepemHL}.

In~fact, there is another equivalent way to obtain the re--expansion of
the strong correction to the $R$--ratio at high energies. Specifically,
instead of following the lines described above, one can expand the
corresponding spectral function~$\rho^{(\ell)}(\sigma)$~(\ref{RhoDef}) and
then perform the integration in Eq.~(\ref{RProp}). In particular, at the
one--loop level the Taylor expansion
of~$\bar\rho^{(1)}_{1}(\sigma)$~(\ref{RhoPert1L})
for~$\sqrt{\sigma}/\Lambda>\exp(\pi/2)$ reads
\begin{equation}
\bar\rho^{(1)}_{1}(\sigma) \simeq \frac{1}{y^2} - \frac{\pi^2}{y^4} +
\co\!\biggl(\frac{1}{y^6}\biggr)\!, \qquad
y=\ln\!\biggl(\!\frac{\sigma}{\Lambda^2}\!\biggr)\!,
\end{equation}
that, after its integration in Eq.~(\ref{RProp}), yields the result
identical to Eq.~(\ref{ATL1L1pUV}). It~is the latter prescription that
will be employed in the next Section for the derivation of an approximate
expression for the $R$--ratio at high energies at an arbitrary loop level.

\section{Re--expansion of the $R$--ratio at high energies: $\pi^2$--terms}
\label{Sect:RepemPi2}

As~outlined in Sect.~\ref{Sect:Repem}, the strong correction to the
$R$--ratio of electron--positron annihilation into hadrons~(\ref{RProp})
can be represented as
\begin{equation}
\label{rYdef}
r(s) =
\int\limits_{s}^{\infty}\!\rho(\sigma)\,\frac{d \sigma}{\sigma} =
\int\limits_{\ln w}^{\infty}\rho_{y}(y)\, d y, \qquad
w=\frac{s}{\Lambda^2},
\end{equation}
where~$y=\ln(\sigma/\Lambda^2)$ and~$\rho_{y}(y) =
\rho\bigl[\Lambda^2\exp(y)\bigr]$ denotes the corresponding spectral
function~(\ref{RhoDef})
\begin{equation}
\label{RhoAppr1}
\rho_{y}(y) = \frac{1}{2 \pi i}\,
\Bigl[d_{y}(y - i \pi) - d_{y}(y + i \pi)\Bigr]\!,
\end{equation}
with Eq.~(\ref{LnCut}) being employed. In~Eq.~(\ref{RhoAppr1}) $d_{y}(y) =
d\bigl[\Lambda^2\exp(y)\bigr]$ stands for the strong correction to the
Adler function being expressed in terms of~$y=\ln(\sigma/\Lambda^2)$.
Applying to the latter the Taylor expansion
\begin{equation}
\label{RhoAppr2}
d_{y}(y \pm i \pi) = d_{y}(y) + \sum_{n=1}^{\infty}\!
\frac{(\pm i\pi)^n}{n !}\frac{d^{n}}{d y^{n}}\, d_{y}(y), \quad
|y| > \pi,
\end{equation}
one can approximate\footnote{It~has to be emphasized here that
Eqs.~(\ref{RhoAppr2}) and~(\ref{RhoAppr3}) are only valid for~$|y|>\pi$,
that eventually bounds the convergence range of the resulting approximate
expression for the $R$--ratio to~$\sqrt{s}/\Lambda > \exp(\pi/2) \simeq
4.81$.} the spectral function~(\ref{RhoAppr1})~by
\begin{equation}
\label{RhoAppr3}
\rho_{y}(y) = - \frac{d}{d y}\, d_{y}(y) -
\sum_{n=1}^{\infty} \frac{(-1)^{n} \pi^{2n}}{(2n+1)!}\,
\frac{d^{2n+1}}{d y^{2n+1}}\, d_{y}(y).
\end{equation}
Therefore, for $\sqrt{s}/\Lambda > \exp(\pi/2) \simeq 4.81$ the strong
correction to the~$R$--ratio~(\ref{rYdef}) acquires the following form
\begin{equation}
\label{rYappr}
r(s) = d(|s|) + \sum_{n=1}^{\infty}
\frac{(-1)^{n} \pi^{2n}}{(2n+1)!}\,
\frac{d^{2n}}{d y^{2n}}\, d_{y}(y) \Biggr|_{y=\ln w}\!.
\end{equation}
In~particular, this equation implies that the strong correction to the
$R$--ratio, being re--expanded at high energies, reproduces the naive
continuation of the Adler function into the timelike domain [the first
term on the right--hand side of Eq.~(\ref{rYappr})] and additionally
produces an infinite number of the so--called $\pi^{2}$--terms.

Then, at the $\ell$--loop level the perturbative expression for the strong
correction to the Adler function reads~(\ref{AdlerPert})
\begin{equation}
\label{dYdef}
d^{(\ell)}_{y}(y) = \sum_{j=1}^{\ell} d_{j}
\left[a^{(\ell)}_{y}(y)\right]^{j}\!,
\end{equation}
where~$a^{(\ell)}_{y}(y)=a\ind{(\ell)}{s}\bigl[\Lambda^2\exp(y)\bigr]$ is
the $\ell$--loop perturbative couplant being expressed in terms
of~$y=\ln(\sigma/\Lambda^2)$. Since the latter satisfies the
renormalization group equation
\begin{equation}
\label{RGEqnY}
\frac{d}{dy}\, a^{(\ell)}_{y}(y) = - \sum_{j=0}^{\ell-1} B_j\!
\left[a^{(\ell)}_{y}(y)\right]^{j+2},
\qquad B_j=\frac{\beta_j}{\beta^{j+1}_{0}},
\end{equation}
the $n$--th derivative of the $j$--th power of the $\ell$--loop couplant
takes the form
\begin{align}
\label{rcYder}
\frac{d^{n}}{dy^{n}}\!\left[a^{(\ell)}_{y}(y)\right]^{j} & \!\! =
(-1)^{n}\!
\sum_{k_{1}=0}^{\ell-1}
\!\ldots\!
\sum_{k_{n}=0}^{\ell-1}\!
\left[a^{(\ell)}_{y}(y)\!\right]^{j+n+k_{1}+\ldots+k_{n}}\!\times
\nonumber \\ & \times\!\!
\left(\prod_{p=1}^{n}B_{k_{p}}\!\!\right)\!\!\!
\left[\prod_{t=0}^{n-1}\Bigl(\!j+t+k_{1}+\ldots+k_{t}\!\Bigr)\!\!\right]\!\!.
\end{align}
Thus, at high energies the $\ell$--loop strong correction to the
$R$--ratio~(\ref{rYdef}) can be approximated~by (see also
Ref.~\cite{Pi2termsHO})
\begin{align}
\label{Pi2TermsGen}
r^{(\ell)}(s) & =
\sum_{j=1}^{\ell} d_{j} \left[a^{(\ell)}_{{\rm s}}(|s|)\right]^{j} -
\sum_{j=1}^{\ell} d_{j}
\sum_{n=1}^{\infty}\! \frac{(-1)^{n+1}\pi^{2n}}{(2n+1)!} \times
\nonumber \\ & \hspace{-7mm} \times
\!\!\sum_{k_{1}=0}^{\ell-1}
\!\ldots\!
\sum_{k_{2n}=0}^{\ell-1}\!\!
\left(\prod_{p=1}^{2n}B_{k_{p}}\!\!\right)\!\!\!
\left[\prod_{t=0}^{2n-1}\!\!
\Bigl(j+t+k_{1}+k_{2}+\ldots+k_{t}\Bigr)\!\!\right]\!\!\! \times
\nonumber \\ & \hspace{-7mm} \times\!\!
\Bigl[a^{(\ell)}_{{\rm s}}(|s|)\Bigr]^{j+2n+k_{1}+k_{2}+\ldots+k_{2n}},
\qquad
\frac{\sqrt{s}}{\Lambda} > \exp\!\biggl(\!\frac{\pi}{2}\!\biggr)\!.
\end{align}

The obtained re--expansion of the strong correction to
the~$R$--ratio~(\ref{Pi2TermsGen}) constitutes the sum of naive
continuation of the strong correction to the Adler function into the
timelike domain~(\ref{RNaive}) and an infinite number of the
$\pi^2$--terms. Equation~(\ref{Pi2TermsGen}) explicitly proves the fact
that at any given loop level the re--expansion of the strong correction to
the $R$--ratio at high energies can be reduced to the form of power series
in the naive continuation of the perturbative couplant into the timelike
domain~$a\ind{(\ell)}{s}(|s|)$. As~one can also note, in the
re--expansion~(\ref{Pi2TermsGen}) the coefficients~$d_j$ corresponding to
various orders of perturbation theory turn out to be all mixed up, i.e.,
the $\ell$--loop contribution to Eq.~(\ref{RProp}) appears to be
re--distributed over the higher--order terms.

\begin{table*}[t]
\caption{Numerical values of the coefficients~$\delta_{j}$~(\ref{rjDef})
embodying the contributions of the
relevant~$\pi^2$--terms~(\ref{Pi2TermsGen}). The~last column employs the
numerical estimation of the Adler function perturbative expansion
coefficient~$d_5$~\cite{RPert5LEstim1}.}
\label{Tab:Delta}
\begin{tabular*}{\textwidth}{@{\extracolsep{\fill}}cccccccc@{}}
\hline
$\nf$ & $\delta_{1}$ & $\delta_{2}$ & $\delta_{3}$ & $\delta_{4}$
      & $\delta_{5}$ & $\delta_{6}$ & $\delta_{7}$ \\
\hline
0 & 0.0000 & 0.0000 & 1.1963 & 5.1127 &    20.455 &     69.081 &    \textit{45.7}\\
1 & 0.0000 & 0.0000 & 1.2735 & 5.4298 &    18.880 &     56.819 &    \textit{7.02}  \\
2 & 0.0000 & 0.0000 & 1.3613 & 5.7583 &    17.118 &     48.532 & $-$\textit{35.7}  \\
3 & 0.0000 & 0.0000 & 1.4622 & 6.0851 &    13.519 &     30.365 & $-$\textit{82.5}  \\
4 & 0.0000 & 0.0000 & 1.5791 & 6.3850 &     6.910 &   $-$3.843 & $-$\textit{115.7} \\
5 & 0.0000 & 0.0000 & 1.7165 & 6.6090 &  $-$3.187 &  $-$45.692 & $-$\textit{83.0}  \\
6 & 0.0000 & 0.0000 & 1.8799 & 6.6638 & $-$21.168 & $-$120.010 &    \textit{142.5} \\
\hline
\end{tabular*}
\end{table*}

As~discussed earlier, if the number of terms retained in
Eq.~(\ref{Pi2TermsGen}) is large enough, then it can provide a rather
accurate approximation of the strong correction to
the~$R$--ratio~(\ref{RProp}) for~$\sqrt{s}/\Lambda > \exp(\pi/2) \simeq
4.81$. However, one usually truncates the
re--expansion~(\ref{Pi2TermsGen}) at the order~$\ell$, thereby neglecting
all the higher--order $\pi^2$--terms\footnote{Though, the latter may not
necessarily be negligible due to a rather large values of the
corresponding coefficients~$\delta_j$~(\ref{rjDef}).}, that results in the
following expression commonly employed in practical applications:
\begin{equation}
\label{RAppr}
R^{(\ell)}_{{\rm appr}}(s) = 1 + r^{(\ell)}_{{\rm appr}}(s),
\quad
r^{(\ell)}_{{\rm appr}}(s) =\!
\sum_{j=1}^{\ell} r_{j} \!\left[a^{(\ell)}_{{\rm s}}(|s|)\right]^{j}\!,
\end{equation}
where
\begin{equation}
\label{rjDef}
r_{j} = d_{j} - \delta_{j}.
\end{equation}
In~this equation $d_j$ stand for the Adler function perturbative expansion
coefficients~(\ref{AdlerPert}), whereas~$\delta_{j}$ embody the
contributions of the relevant~$\pi^2$--terms.

Equation~(\ref{Pi2TermsGen}) implies that the $\pi^2$--terms do not
appear in the first and second orders of perturbation theory, namely,
\begin{equation}
\label{Delta1and2}
\delta_{1} = 0, \qquad \delta_{2} = 0,
\end{equation}
that makes~$R^{(\ell)}_{{\rm appr}}(s)$~(\ref{RAppr}) identical to the
naive expression~$R^{(\ell)}_{{\rm naive}}(s)$~(\ref{RNaive}) for~$\ell=1$
and~$\ell=2$. Starting from the third order of perturbation theory (i.e.,
for~$\ell \ge 3$) the coefficients~$\delta_j$~(\ref{rjDef}) are no longer
vanishing and constitute a combination of the pertinent perturbative
expansion coefficients~$d_{j}$ and~$\beta_{j}$ of the first~$(\ell-2)$
orders. Specifically, the third--order and fourth--order coefficients
read~\cite{Pi2Terms1, RPert5LEstim1, ProsperiAlpha}
\begin{equation}
\label{Delta3}
\delta_{3} = \frac{\pi^2}{3} d_{1} B_{0}^{2} = \frac{\pi^2}{3} d_{1}
\end{equation}
and
\begin{equation}
\label{Delta4}
\delta_{4} = \frac{\pi^2}{3}\!\left(\frac{5}{2} d_{1} B_{0}B_{1} +
3 d_{2}B_{0}^{2}\!\right)\! = \frac{\pi^2}{3}\!\left(\frac{5}{2} d_{1} B_{1} +
3 d_{2}\!\right)\!\!,
\end{equation}
respectively. In~these equations~$d_j$ denote the Adler function
perturbative expansion coefficients~(\ref{AdlerPert}), whereas
$B_{j}=\beta_j/\beta^{j+1}_{0}$ stands for the combination of perturbative
coefficients of the renormalization group~$\beta$~function. In~turn, at
the fifth and sixth orders the coefficients~$\delta_j$~(\ref{rjDef}) can
be represented~as~\cite{RPert5LEstim1, Book, Pi2termsHO}
\begin{align}
\label{Delta5}
\delta_{5} &=
\frac{\pi^2}{3} \biggl[ \frac{3}{2} d_{1}\! \Bigl( B_{1}^{2} + 2 B_{2} \!\Bigr)
+ 7 d_{2} B_{1} + 6 d_3 \biggr]\!
- \frac{\pi^4}{5} d_{1}, \\[2mm]
\label{Delta6}
\delta_{6} &= \frac{\pi^2}{3}
\biggl[ \frac{7}{2} d_{1}\! \Bigl( B_{1} B_{2} + B_{3}\! \Bigr)
+ 4 d_{2} \left( B_{1}^{2} + 2 B_{2} \right) +
\nonumber \\[1mm] &
+ \frac{27}{2} d_3 B_{1} + 10 d_{4} \biggr]\! -
\frac{\pi^4}{5}\! \left( \frac{77}{12} d_{1} B_{1} + 5 d_{2}\! \right)\!\!.
\end{align}
The seventh-- and eighth--order coefficients~$\delta_j$~(\ref{rjDef}) have
recently been calculated as well, specifically~\cite{Book, Pi2termsHO}
\begin{align}
\label{Delta7}
\delta_{7} & =
\frac{\pi^2}{3} \Biggl[
4 d_{1} \!\left(\! B_{1} B_{3} + \frac{1}{2} B_{2}^{2} + B_{4} \!\right) +
9 d_{2} \Bigl(\! B_{1} B_{2} + B_{3} \!\Bigr) +
\nonumber \\[1.25mm]
& + \frac{15}{2} d_{3} \Bigl(\! B_{1}^{2} + 2 B_{2} \!\Bigr) +
22 d_{4} B_{1} + 15 d_{5} \Biggr] -
\nonumber \\[1.25mm]
& - \frac{\pi^4}{5} \left[
\frac{5}{6} d_{1} \Bigl(\! 17 B_{1}^{2} + 12 B_{2} \!\Bigr)
+ \frac{57}{2} d_{2} B_{1} + 15 d_{3} \right]\! + \frac{\pi^6}{7} d_1
\end{align}
and
\begin{align}
\label{Delta8}
\delta_{8} & =
\frac{\pi^2}{3} \Biggl[
\frac{9}{2} d_{1} \Bigl(\! B_{1} B_{4} + B_{2} B_{3} + B_{5} \!\Bigr) +
\nonumber \\ & +
10 d_{2} \!\left(\! B_{1} B_{3} + \frac{1}{2} B_{2}^{2} + B_{4}\! \right)
+ \frac{33}{2} d_{3} \Bigl(\! B_{1} B_{2} + B_{3} \!\Bigr) +
\nonumber \\
& + 12 d_{4} \Bigl(\! B_{1}^{2} + 2 B_{2} \!\Bigr) +
\frac{65}{2} d_{5} B_{1} + 21 d_{6} \Biggr] -
\nonumber \\
& - \frac{\pi^4}{5} \Biggl[
\frac{15}{8} d_{1} \!\Bigl(\! 7 B_{1}^{3} + 22 B_{1} B_{2} + 8 B_{3} \!\Bigr) +
\nonumber \\ & +
\frac{5}{12} d_{2} \Bigl(\! 139 B_{1}^{2} + 96 B_{2} \!\Bigr)
+ \frac{319}{4} d_{3} B_{1} + 35 d_{4} \Biggr] +
\nonumber \\
& + \frac{\pi^6}{7} \left( \frac{223}{20} d_{1} B_{1} + 7 d_{2} \right)\!.
\end{align}
The~explicit expressions for the coefficients~$\delta_j$~(\ref{rjDef}) at
the higher orders can be found in App.~C of Ref.~\cite{Book}.

\begin{table*}[t]
\caption{Numerical values of the coefficients~$r_{j}$ of the re--expanded
approximate $R$--ratio~(\ref{RAppr}). The~last column employs the
numerical estimation of the Adler function perturbative expansion
coefficient~$d_5$~\cite{RPert5LEstim1}.}
\label{Tab:RPert}
\begin{tabular*}{\textwidth}{@{\extracolsep{\fill}}cccccc@{\extracolsep{\fill}}}
\hline
$\nf$ & $r_{1}=d_{1}$ & $r_{2}=d_{2}$ & $r_{3}=d_{3}-\delta_{3}$
      & $r_{4}=d_{4}-\delta_{4}$ & $r_{5}=d_{5}-\delta_{5}$ \\
\hline
0 & 0.3636 & 0.2626 & $-$0.3191 & $-$2.7383 & $-$\textit{15.1} \\
1 & 0.3871 & 0.2803 & $-$0.4788 & $-$3.2413 & $-$\textit{14.2} \\
2 & 0.4138 & 0.3005 & $-$0.6476 & $-$3.6116 & $-$\textit{13.4} \\
3 & 0.4444 & 0.3239 & $-$0.9028 & $-$4.1703 & $-$\textit{11.0} \\
4 & 0.4800 & 0.3513 & $-$1.2923 & $-$5.0409 & $-$\textit{5.75} \\
5 & 0.5217 & 0.3836 & $-$1.8186 & $-$5.9601 &    \textit{3.21} \\
6 & 0.5714 & 0.4225 & $-$2.6630 & $-$7.5590 &    \textit{21.4} \\
\hline
\end{tabular*}
\end{table*}

\begin{table*}[t]
\caption{The~relative weight $(1+|d_{j}/\delta_{j}|)^{-1}\!\times\!100\%$
of the~$\pi^2$--terms in the coefficients~$r_j$ of the re--expanded
approximate $R$--ratio~(\ref{RAppr}). The~last column employs the
numerical estimation of the Adler function perturbative expansion
coefficient~$d_5$~\cite{RPert5LEstim1}.}
\label{Tab:DeltaToD}
\begin{tabular*}{\textwidth}{@{\extracolsep{\fill}}cccccc@{\extracolsep{\fill}}}
\hline
$\nf$ & $j=1$ & $j=2$ & $j=3$ & $j=4$ & $j=5$ \\
\hline
0 & $0.00\,\%$ & $0.00\,\%$ & $57.7\,\%$ & $68.3\,\%$ & \textit{79.1}$\,\%$ \\
1 & $0.00\,\%$ & $0.00\,\%$ & $61.6\,\%$ & $71.3\,\%$ & \textit{80.1}$\,\%$ \\
2 & $0.00\,\%$ & $0.00\,\%$ & $65.6\,\%$ & $72.8\,\%$ & \textit{82.1}$\,\%$ \\
3 & $0.00\,\%$ & $0.00\,\%$ & $72.3\,\%$ & $76.1\,\%$ & \textit{84.3}$\,\%$ \\
4 & $0.00\,\%$ & $0.00\,\%$ & $84.6\,\%$ & $82.6\,\%$ & \textit{85.6}$\,\%$ \\
5 & $0.00\,\%$ & $0.00\,\%$ & $94.4\,\%$ & $91.1\,\%$ & \textit{99.2}$\,\%$ \\
6 & $0.00\,\%$ & $0.00\,\%$ & $70.6\,\%$ & $88.2\,\%$ & \textit{98.8}$\,\%$ \\
\hline
\end{tabular*}
\end{table*}

Table~\ref{Tab:Delta} presents the numerical values of the first seven
coefficients $\delta_j$~(\ref{rjDef}), which embody the contributions of
the relevant~$\pi^2$--terms~(\ref{Pi2TermsGen}).
Tables~\ref{Tab:AdlerPert} and~\ref{Tab:Delta} make it evident that in
Eq.~(\ref{RAppr}) the coefficients~$\delta_j$ can in no way be regarded as
small corrections to the Adler function perturbative expansion
coefficients~$d_j$~(\ref{AdlerPert}) for~$j \ge 3$. On~the contrary, the
values of coefficients~$\delta_j$ significantly exceed the values of
respective perturbative coefficients~$d_j$, thereby constituting the
dominant contribution to the coefficients~$r_j$~(\ref{rjDef}), see also
Tabs.~\ref{Tab:RPert} and~\ref{Tab:DeltaToD}. As~will be discussed below,
eventually this results in an essential distortion of the re--expanded
approximation~$R^{(\ell)}_{\text{appr}}(s)$~(\ref{RAppr}) with respect to
the naive expression~$R^{(\ell)}_{\text{naive}}(s)$~(\ref{RNaive}).
In~particular, the higher--order terms
of~$R^{(\ell)}_{\text{appr}}(s)$~(\ref{RAppr}) turn out to be
substantially amplified and even sign--reversed with respect to those
of~$R^{(\ell)}_{\text{naive}}(s)$~(\ref{RNaive}), see
Tabs.~\ref{Tab:AdlerPert} and~\ref{Tab:RPert}. It~is worthwhile to note
also that, as one can infer from Tabs.~\ref{Tab:Delta}
and~\ref{Tab:DeltaToD}, the values of coefficients~$\delta_j$ rapidly
increase as the order~$j$ increases, that makes the loop convergence
of~$R\ind{(\ell)}{appr}(s)$~(\ref{RAppr}) worse than that of
both~$R^{(\ell)}(s)$~(\ref{RProp})
and~$R^{(\ell)}_{\text{naive}}(s)$~(\ref{RNaive}), see
Sect.~\ref{Sect:RepemHL} for the details.

\section{Spectral function at the higher--loop levels}
\label{Sect:RhoHL}

It is certainly desirable to leave the truncated re--expanded
approximation $R\ind{(\ell)}{appr}(s)$~(\ref{RAppr}) aside and have the
function~$R^{(\ell)}(s)$~(\ref{RProp}) calculated in a straightforward way
beyond the one--loop level. To~achieve this objective, the corresponding
explicit expression for the involved spectral
function~$\rho^{(\ell)}(\sigma)$~(\ref{RhoDef}) is required. Despite the
latter becomes rather cumbrous for $\ell \ge 2$, the following method
enables one to calculate~$\rho^{(\ell)}(\sigma)$ explicitly at an
arbitrary\footnote{It is assumed that the involved perturbative
coefficients~$d_j$ and~$\beta_j$ are known.} loop level.

Specifically, it proves to be convenient to express the spectral
function~$\rho^{(\ell)}(\sigma)$~(\ref{RhoDef}) in terms of the
so--called ``partial'' spectral functions~$\bar\rho^{(\ell)}_{j}(\sigma)$
corresponding to the \mbox{$j$--th} power of the $\ell$--loop perturbative
couplant~(\ref{AItGen}), namely
\begin{equation}
\label{RhoPert2}
\rho^{(\ell)}(\sigma) = \sum_{j=1}^{\ell} d_{j}\,
\bar\rho^{(\ell)}_{j}(\sigma),
\end{equation}
where
\begin{equation}
\label{RhoPart}
\bar\rho^{(\ell)}_{j}(\sigma) = \frac{1}{2\pi i}
\lim_{\varepsilon \to 0_{+}}\! \left\{\!\!
\left[a\ind{(\ell)}{s}\!(-\sigma-i\varepsilon)\right]^{j} \!\!-\!
\left[a\ind{(\ell)}{s}\!(-\sigma+i\varepsilon)\right]^{j}\!
\right\}\!\!.
\end{equation}
To~obtain the partial spectral function
$\bar\rho^{(\ell)}_{j}(\sigma)$~(\ref{RhoPart}) for any $j \ge 1$ it
appears to be enough to calculate only the real and imaginary parts of the
$\ell$--loop couplant $a\ind{(\ell)}{s}(Q^2)$ at the edges of its cut:
\begin{equation}
\label{AReImDef}
\lim_{\varepsilon \to 0_{+}}
a\ind{(\ell)}{s}(-\sigma \pm i\varepsilon) =
\ARe{(\ell)}(\sigma) \mp i \pi \AIm{(\ell)}(\sigma).
\end{equation}
Here~$\ARe{(\ell)}(\sigma)$ and~$\AIm{(\ell)}(\sigma)$ are the real
functions of their arguments and~$\sigma \ge 0$ is assumed. For~positive
integer values of~$j$ the following equation holds
\begin{align}
\label{RhoPartAux1}
\lim_{\varepsilon \to 0_{+}}
\Bigl[a\ind{(\ell)}{s}(-\sigma \pm i\varepsilon)\Bigr]^{j} & =
\sum_{k=0}^{j} \binom{j}{k}\, (\mp i\pi)^{k} \times
\nonumber \\
& \times \Bigl[\ARe{(\ell)}(\sigma)\Bigr]^{j-k}\,
\Bigl[\AIm{(\ell)}(\sigma)\Bigr]^{k},
\end{align}
with
\begin{equation}
\label{BinomDef}
\binom{n}{m} = \frac{n!}{m!\,(n-m)!}
\end{equation}
being the binomial coefficient. Then, it is convenient to isolate in
Eq.~(\ref{RhoPartAux1}) its real and imaginary parts, specifically
\begin{align}
\lim_{\varepsilon \to 0_{+}}
\Bigl[a\ind{(\ell)}{s}(-\sigma \pm i\varepsilon)\Bigr]^{j} & =
\sum_{k=0}^{K(j+1)} \binom{j}{2k}\, (-1)^{k}(\pi)^{2k} \times
\nonumber \\
& \times
\Bigl[\ARe{(\ell)}(\sigma)\Bigr]^{j-2k}\,
\Bigl[\AIm{(\ell)}(\sigma)\Bigr]^{2k} \mp \quad \nonumber \\[1.25mm]
& \mp i \pi
\sum_{k=0}^{K(j)} \binom{j}{2k+1}\, (-1)^{k}(\pi)^{2k} \times
\nonumber \\
& \times
\Bigl[\ARe{(\ell)}(\sigma)\Bigr]^{j-2k-1}\,
\Bigl[\AIm{(\ell)}(\sigma)\Bigr]^{2k+1},
\end{align}
where
\begin{equation}
\label{KImDef}
K(j) = \frac{j-2}{2} + \frac{j \;\mbox{mod}\; 2}{2}
\end{equation}
and $(j \;\mbox{mod}\; n)$ denotes the remainder on division of~$j$
by~$n$. Therefore, the partial spectral function~(\ref{RhoPart}) reads
(see also Refs.~\cite{Review, QCDDPT1})
\begin{align}
\label{RhoPartHO}
\bar\rho^{(\ell)}_{j}(\sigma) & = \sum_{k=0}^{K(j)}
\binom{j}{2k+1}\, (-1)^{k}\, \pi^{2k} \times
\nonumber \\
& \times
\Bigl[\ARe{(\ell)}(\sigma)\Bigr]^{j-2k-1}\,
\Bigl[\AIm{(\ell)}(\sigma)\Bigr]^{2k+1},
\end{align}
with $\sigma \ge 0$ and $j \ge 1$ being assumed. In~particular, the first
five relations~(\ref{RhoPartHO}) acquire a compact form
\begin{align}
\bar\rho^{(\ell)}_{1}(\sigma) & = \AIm{(\ell)}(\sigma),
%\end{align}
\\[1.5mm]
%\begin{align}
\bar\rho^{(\ell)}_{2}(\sigma) & = 2\, \AIm{(\ell)}(\sigma)\,
     \ARe{(\ell)}(\sigma),
%\end{align}
\\[1.5mm]
%\begin{align}
\bar\rho^{(\ell)}_{3}(\sigma) & = \AIm{(\ell)}(\sigma)
     \left\{\!3\Bigl[\ARe{(\ell)}(\sigma)\Bigr]^{2} \!-
     \pi^{2}\Bigl[\AIm{(\ell)}(\sigma)\Bigr]^{2}\right\},
%\end{align}
\\[1.5mm]
%\begin{align}
\bar\rho^{(\ell)}_{4}(\sigma) & = 4\, \AIm{(\ell)}(\sigma)\,
     \ARe{(\ell)}(\sigma)
     \left\{\!\Bigl[\ARe{(\ell)}(\sigma)\Bigr]^{2} \!-
     \pi^{2}\Bigl[\AIm{(\ell)}(\sigma)\Bigr]^{2}\right\},
%\end{align}
\\[1.5mm]
%\begin{align}
\bar\rho^{(\ell)}_{5}(\sigma) & = \AIm{(\ell)}(\sigma)
     \biggl\{\!5\Bigl[\ARe{(\ell)}(\sigma)\Bigr]^{4} \!-
     10 \pi^{2}\Bigl[\AIm{(\ell)}(\sigma)\ARe{(\ell)}(\sigma)\Bigr]^{2}
\nonumber \\
     & + \pi^{4}\Bigl[\AIm{(\ell)}(\sigma)\Bigr]^{4}\biggr\}.
\end{align}

In turn, the $\ell$--loop functions~$\ARe{(\ell)}(\sigma)$
and~$\AIm{(\ell)}(\sigma)$ entering Eq.~(\ref{RhoPartHO}) can also be
explicitly calculated in a similar~way. Specifically, as mentioned in
Sect.~\ref{Sect:Repem}, the perturbative QCD~couplant
$a\ind{(\ell)}{s}(Q^2)$ can be represented as the double
sum~(\ref{AItGen}) comprised of the functions
\begin{equation}
\label{RCLnm}
\bar a_{n}^{m}(Q^2) = \frac{\ln^{m}(\ln z)}{\ln^n z},
\qquad z = \frac{Q^2}{\Lambda^2}.
\end{equation}
It is worthwhile to decompose the function~$\bar
a_{n}^{m}(Q^2)$~(\ref{RCLnm}) at the edges of its cut into the real and
imaginary parts, namely
\begin{equation}
\label{UnmVnm}
\lim_{\varepsilon \to 0_{+}} \bar a_{n}^{m}(-\sigma \pm i\varepsilon) =
u_{n}^{m}(\sigma) \mp i \pi v_{n}^{m}(\sigma),
\end{equation}
where $u_{n}^{m}(\sigma)$ and~$v_{n}^{m}(\sigma)$ are the real functions
of their arguments and $\sigma \ge 0$ is~assumed. Therefore, the functions
$\ARe{(\ell)}(\sigma)$ and~$\AIm{(\ell)}(\sigma)$~(\ref{AReImDef}) take
the following form
\begin{align}
\label{AReSum}
\ARe{(\ell)}(\sigma) & =
\sum_{n=1}^{\ell}\sum_{m=0}^{n-1} b^{m}_{n}\, u_{n}^{m}(\sigma),
\\
\label{AImSum}
\AIm{(\ell)}(\sigma) & =
\sum_{n=1}^{\ell}\sum_{m=0}^{n-1} b^{m}_{n}\, v_{n}^{m}(\sigma).
\end{align}
On the left--hand side of Eq.~(\ref{RCLnm}) and in
Eqs.~(\ref{UnmVnm})--(\ref{AImSum}) the integer superscripts~$m$ are not
to be confused with respective powers, whereas the
coefficients~$b^{m}_{n}$ have been specified in Eq.~(\ref{AItGen}).

\begin{figure*}[t]
\centering%
{\includegraphics[width=70mm,clip]{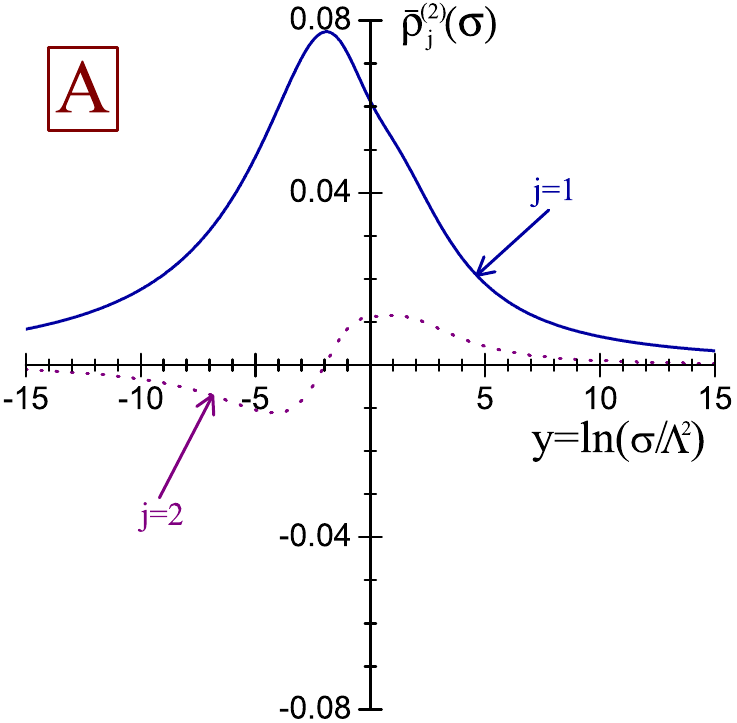}%
\hspace{7.0mm}%
\includegraphics[width=70mm,clip]{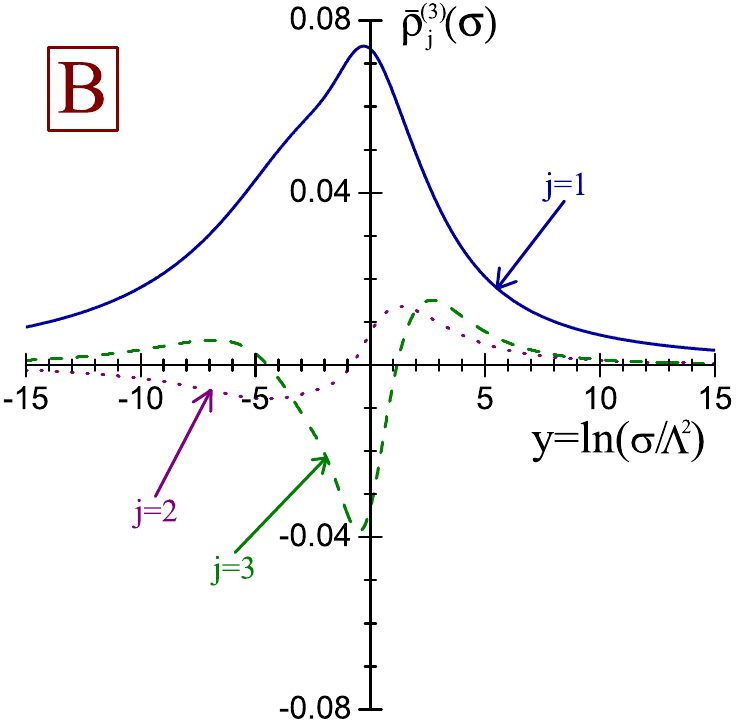}}
\vskip7.0mm
{\includegraphics[width=70mm,clip]{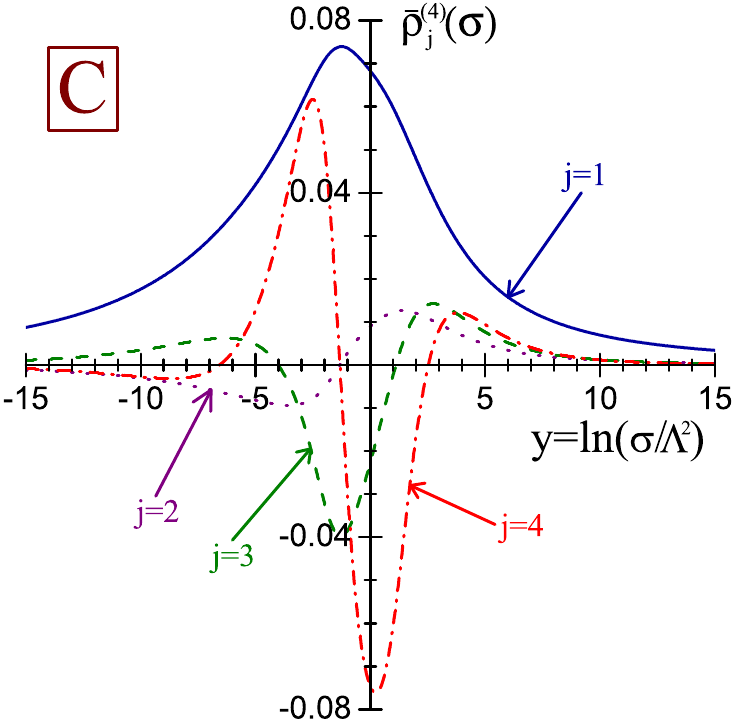}%
\hspace{7.0mm}%
\includegraphics[width=70mm,clip]{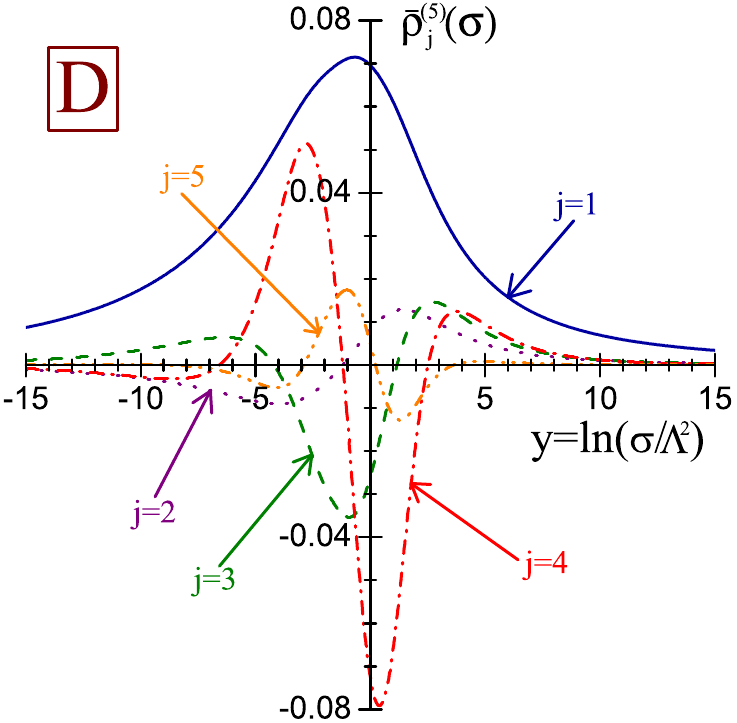}}
\caption{The partial spectral functions
$\bar\rho^{(\ell)}_{j}(\sigma)$~(\ref{RhoPartHO}) corresponding to the
\mbox{$j$--th} power ($1 \le j \le \ell$) of the $\ell$--loop ($2 \le \ell
\le 5$) perturbative QCD couplant~$a^{(\ell)}_{{\rm s}}(Q^2)$.
Plot~A:~two--loop level ($\ell=2$, $1 \le j \le 2$). Plot~B:~three--loop
level ($\ell=3$, $1 \le j \le 3$), the
function~$\bar\rho^{(3)}_{3}(\sigma)$ is scaled by the factor of~$10$.
Plot~C:~four--loop level ($\ell=4$, $1 \le j \le 4$), the
functions~$\bar\rho^{(4)}_{3}(\sigma)$ and~$\bar\rho^{(4)}_{4}(\sigma)$
are scaled by the factors of~$10$ and~$10^2$, respectively.
Plot~D:~five--loop level ($\ell=5$, $1 \le j \le 5$), the
functions~$\bar\rho^{(5)}_{3}(\sigma)$, $\bar\rho^{(5)}_{4}(\sigma)$,
and~$\bar\rho^{(5)}_{5}(\sigma)$ are scaled by the factors of~$10$,
$10^2$, and~$10^2$, respectively.}
\label{Plot:Rho5L}
\end{figure*}

Then, to calculate the functions~$u_{n}^{m}(\sigma)$
and~$v_{n}^{m}(\sigma)$ entering Eqs.~(\ref{AReSum}) and~(\ref{AImSum}),
it is convenient to split the left--hand side of Eq.~(\ref{UnmVnm}) into
two factors:
\begin{equation}
\label{AnmF}
\lim_{\varepsilon \to 0_{+}} \bar a_{n}^{m}(-\sigma \pm i\varepsilon) =
\lim_{\varepsilon \to 0_{+}}
\Bigl[\bar a_{n}^{0}(-\sigma \pm i\varepsilon)\,
\bar a_{0}^{m}(-\sigma \pm i\varepsilon)\!\Bigr]\!.
\end{equation}
The first factor on the right--hand side of Eq.~(\ref{AnmF}) reads
\begin{equation}
\lim_{\varepsilon \to 0_{+}} \bar a_{n}^{0}(-\sigma \pm i\varepsilon) =
\frac{(y \mp i\pi)^{n}}{(y^2 + \pi^2)^{n}},
\qquad y=\ln\biggl(\!\frac{\sigma}{\Lambda^2}\!\biggr)\!.
\end{equation}
Proceeding along the same lines as earlier, one can cast the numerator on
the right--hand side of this equation to
\begin{align}
(y \mp i\pi)^{n} & = \!\!\!
\sum_{k=0}^{K(n+1)} \!\binom{n}{2k} (-1)^{k} \pi^{2k} y^{n-2k} \mp
\nonumber \\
& \mp
i\pi \!\! \sum_{k=0}^{K(n)} \!\binom{n}{2k+1} (-1)^{k} \pi^{2k} y^{n-2k-1},
\end{align}
with~$K(n)$ being specified in Eq.~(\ref{KImDef}). Hence, the
functions~$u_{n}^{m}(\sigma)$ and~$v_{n}^{m}(\sigma)$~(\ref{UnmVnm})
for~$m=0$ read
\begin{align}
\label{Un0Def}
u_{n}^{0}(\sigma) & = \frac{1}{(y^2 + \pi^2)^{n}}
\sum_{k=0}^{K(n+1)} \!\binom{n}{2k} (-1)^{k} \pi^{2k} y^{n-2k},
\\
\label{Vn0Def}
v_{n}^{0}(\sigma) & = \frac{1}{(y^2 + \pi^2)^{n}}
\sum_{k=0}^{K(n)} \!\binom{n}{2k+1} (-1)^{k} \pi^{2k} y^{n-2k-1},
\end{align}
with $n \ge 1$ being assumed.

In turn, for~$m \ge 1$ the second factor on the right--hand side of
Eq.~(\ref{AnmF}) takes the following form
\begin{equation}
\label{AnmAux1}
\lim_{\varepsilon \to 0_{+}} \bar a_{0}^{m}(-\sigma \pm i\varepsilon) =
\Bigl[\ln(y \pm i\pi)\Bigr]^{m},
\end{equation}
where~$y=\ln(\sigma/\Lambda^2)$. Since for real~$a$ and~$b$
\begin{equation}
\label{CompLn}
\ln (a \pm i b) = \ln\!\sqrt{\! a^2 + b^2} \pm
i\pi\!\!\left[\frac{1}{2} - \frac{1}{\pi}\arctan\!\left(\frac{a}{b}\right)\!\right]\!\!,
\quad b>0,
\end{equation}
Eq.~(\ref{AnmAux1}) can be rewritten as
\begin{equation}
\label{LLm}
\lim_{\varepsilon \to 0_{+}} \bar a_{0}^{m}(-\sigma \pm i\varepsilon) =
\sum\limits_{k=0}^{m}\binom{m}{k}
(\pm i\pi)^{k} \Bigl[L_{1}(y)\Bigr]^{m-k}\, \Bigl[L_{2}(y)\Bigr]^{k},
\end{equation}
where
\begin{equation}
\label{L12Def}
L_{1}(y) = \ln\!\sqrt{y^{2}+\pi^{2}}, \qquad
L_{2}(y) = \frac{1}{2} - \frac{1}{\pi}\arctan\!\left(\frac{y}{\pi}\right)\!.
\end{equation}
Following the same steps as above, one can cast Eq.~(\ref{LLm}) to
\begin{equation}
\lim_{\varepsilon \to 0_{+}} \bar a_{0}^{m}(-\sigma \pm i\varepsilon) =
u_{0}^{m}(\sigma) \mp i \pi v_{0}^{m}(\sigma),
\end{equation}
where
\begin{align}
\label{U0mDef}
u_{0}^{m}(\sigma) & = \!\sum\limits_{k=0}^{K(m+1)}\binom{m}{2k}
(-1)^{k}\pi^{2k} \times
\nonumber \\
& \times \Bigl[L_{1}(y)\Bigr]^{m-2k}\, \Bigl[L_{2}(y)\Bigr]^{2k},
\\[1.75mm]
\label{V0mDef}
v_{0}^{m}(\sigma) & = \!\sum\limits_{k=0}^{K(m)}\binom{m}{2k+1}
(-1)^{k+1}\pi^{2k} \times
\nonumber \\
& \times \Bigl[L_{1}(y)\Bigr]^{m-2k-1}\, \Bigl[L_{2}(y)\Bigr]^{2k+1},
\end{align}
$K(m)$ is defined in Eq.~(\ref{KImDef}), and $m \ge 1$ is assumed.

\begin{figure}[t]
\centering
\includegraphics[width=72.5mm,clip]{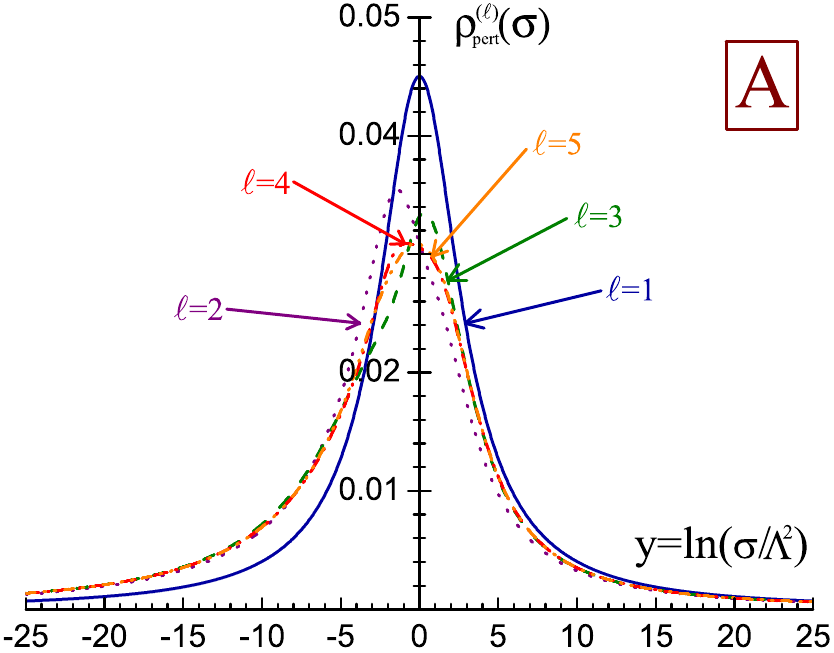}%
\hspace{7.5mm}%
\raisebox{2.25mm}{\includegraphics[width=72.5mm,clip]{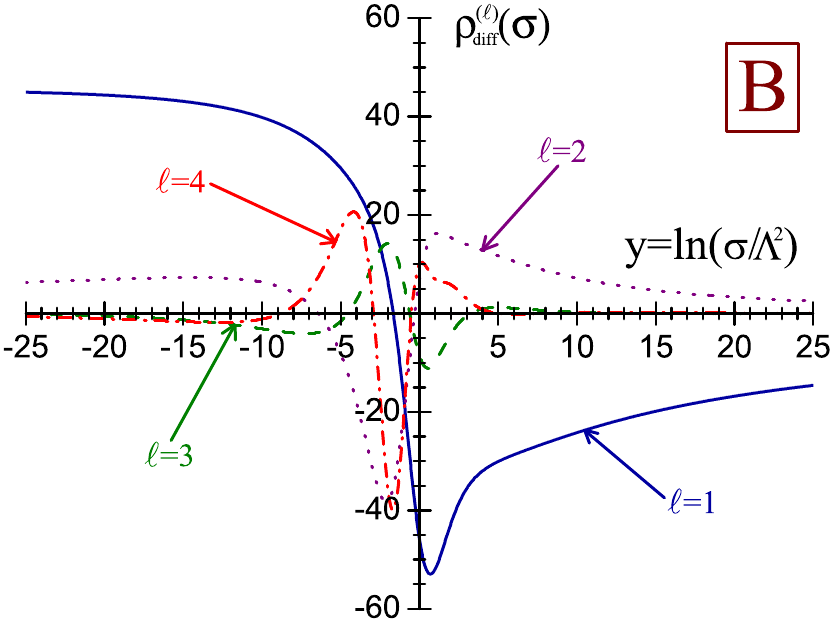}}
\caption{Plot~A: The spectral function
$\rho^{(\ell)}(\sigma)$~(\ref{RhoPertHO}) at the first five loop levels ($1
\le \ell \le 5$). Plot~B: The relative
difference~$\rho^{(\ell)}_{\text{diff}}(\sigma)$~(\ref{RhoDiff}) between
the $\ell$--loop and $(\ell+1)$--loop spectral functions~(\ref{RhoPertHO})
at various loop levels. The function~$\rho^{(4)}_{\text{diff}}(\sigma)$ is
scaled by the factor of~$10$.}
\label{Plot:RhoPert}
\end{figure}

Therefore, the functions~$u_{n}^{m}(\sigma)$
and~$v_{n}^{m}(\sigma)$~(\ref{UnmVnm}) read
\begin{equation}
\label{Unm}
u_{n}^{m}(\sigma) =
\begin{cases}
u_{n}^{0}(\sigma),& \text{if $\, m=0$},\\[1.25mm]
u_{n}^{0}(\sigma) u_{0}^{m}(\sigma) -
\pi^2 v_{n}^{0}(\sigma) v_{0}^{m}(\sigma),\quad& \text{if $\, m \ge 1$},
\end{cases}
\end{equation}
and
\begin{equation}
\label{Vnm}
v_{n}^{m}(\sigma) =
\begin{cases}
v_{n}^{0}(\sigma),& \text{if $\, m=0$},\\[1.25mm]
v_{n}^{0}(\sigma) u_{0}^{m}(\sigma) +
u_{n}^{0}(\sigma) v_{0}^{m}(\sigma),\quad& \text{if $\, m \ge 1$},
\end{cases}
\end{equation}
with $n \ge 1$ being assumed. Thus, the explicit expression for
the~$\ell$--loop spectral
function~$\rho^{(\ell)}(\sigma)$~(\ref{RhoPert2}) takes the following
form:
\begin{align}
\label{RhoPertHO}
\rho^{(\ell)}(\sigma) & =
\sum_{j=1}^{\ell} d_{j} \sum_{k=0}^{K(j)}
\binom{j}{2k+1} (-1)^{k}\, \pi^{2k} \times
\nonumber \\
& \times\!
\Biggl[\sum_{n=1}^{\ell}\sum_{m=0}^{n-1} b^{m}_{n}\, u_{n}^{m}(\sigma)\Biggr]^{j-2k-1} \times
\nonumber \\
& \times\!
\Biggl[\sum_{n=1}^{\ell}\sum_{m=0}^{n-1} b^{m}_{n}\, v_{n}^{m}(\sigma)\Biggr]^{2k+1},
\end{align}
where the functions~$u_{n}^{m}(\sigma)$ and~$v_{n}^{m}(\sigma)$ are
specified in Eqs.~(\ref{Unm}) and~(\ref{Vnm}), respectively.

The higher--loop partial spectral
functions~$\bar\rho^{(\ell)}_{j}(\sigma)$~(\ref{RhoPartHO}), which
correspond to the \mbox{$j$--th} power ($1 \le j \le \ell$) of the
$\ell$--loop ($2 \le \ell \le 5$) perturbative QCD
couplant~$a\ind{(\ell)}{s}(Q^2)$, are displayed in~Fig.~\ref{Plot:Rho5L}.
As~one can infer from this Figure, the
functions~$\bar\rho^{(\ell)}_{j}(\sigma)$~(\ref{RhoPartHO}) vanish at
both~$\sigma\to\infty$ and~$\sigma \to 0$, the higher--order
functions~$\bar\rho^{(\ell)}_{j}(\sigma)$ ($j \ge 2$) being substantially
suppressed with respect to those of the preceding orders. For~example,
Fig.~\ref{Plot:Rho5L}$\,$D implies that at the five--loop level the
maximum value of the fifth--order function~$\bar\rho^{(5)}_{5}(\sigma)$ is
about three orders of magnitude less than the maximum value of the
first--order function~$\bar\rho^{(5)}_{1}(\sigma)$. In~turn, as it will be
discussed in the next Section, the fact that the
function~$\bar\rho^{(\ell)}_{j+1}(\sigma)$ is subdominant
to~$\bar\rho^{(\ell)}_{j}(\sigma)$ eventually results in an enhanced
higher--loop stability of the proper expression for the
$R$--ratio~(\ref{RProp}) at moderate and low energies with respect to both
its naive form~(\ref{RNaive}) and the commonly employed truncated
re--expanded approximation~(\ref{RAppr}).

The plots of the spectral
function~$\rho^{(\ell)}(\sigma)$~(\ref{RhoPertHO}) at the first five loop
levels ($1 \le \ell \le 5$) are displayed in Fig.~\ref{Plot:RhoPert}$\,$A.
As~one can infer from this Figure, the function~$\rho^{(\ell)}(\sigma)$
vanishes at both~$\sigma\to\infty$ and~$\sigma \to 0$. Specifically, at
the higher--loop levels ($\ell \ge 2$)
\begin{equation}
\rho^{(\ell)}(\sigma) \simeq \frac{d_{1}}{y^2} +
\co\left(\frac{1}{y^3}\right)\!,
\quad y=\ln\biggl(\frac{\sigma}{\Lambda^2}\biggr)\!,
\quad y\to\infty
\end{equation}
and
\begin{equation}
\rho^{(\ell)}(\sigma) \simeq \frac{d_{1}(1+B_{1})}{y^2} +
\co\left(\frac{1}{y^3}\right)\!, \qquad y \to -\infty.
\end{equation}
Additionally, Fig.~\ref{Plot:RhoPert}$\,$A implies that the range of~$y$,
where the difference between~$\rho^{(\ell)}(\sigma)$
and~$\rho^{(\ell+1)}(\sigma)$ is sizable, is located in the vicinity
of~$y=0$ and becomes smaller at larger~$\ell$. This issue is also
elucidated by Fig.~\ref{Plot:RhoPert}$\,$B, which shows the relative
difference between the $\ell$--loop and $(\ell+1)$--loop spectral
functions~(\ref{RhoPertHO})
\begin{equation}
\label{RhoDiff}
\rho\ind{(\ell)}{diff}(\sigma) =
\!\left[1-\frac{\rho^{(\ell)}(\sigma)}
{\rho^{(\ell+1)}(\sigma)}\right]\!\!\times 100\%
\end{equation}
at various loop levels.

\section{$R$--ratio at the higher--loop levels}
\label{Sect:RepemHL}

The obtained in the previous Section explicit expression for the
$\ell$--loop spectral function~$\rho^{(\ell)}(\sigma)$~(\ref{RhoPertHO})
enables one to calculate the $R$--ratio of electron--positron annihilation
into hadrons~(\ref{RProp}) at an arbitrary loop level (assuming that the
involved perturbative coefficients~$\beta_j$ and~$d_j$ are available).
The~integration in Eq.~(\ref{RProp}) can be directly performed for the
function~$\rho^{(\ell)}(\sigma)$~(\ref{RhoPertHO}) as a whole, though, for
the illustrative purposes it is somewhat convenient to keep the
contributions of the partial spectral functions
$\bar\rho^{(\ell)}_{j}(\sigma)$~(\ref{RhoPartHO}) to Eq.~(\ref{RhoPert2})
separately from each other, that casts Eq.~(\ref{RProp})~to
\begin{equation}
\label{RProp2}
R^{(\ell)}(s) = 1 + r^{(\ell)}(s), \qquad
r^{(\ell)}(s) = \sum\limits_{j=1}^{\ell} d_{j}\,\ATL{(\ell)}{j}(s).
\end{equation}
In~this equation
\begin{equation}
\label{ATLDef}
\ATL{(\ell)}{j}(s) = \int\limits_{s}^{\infty}\!\bar\rho^{(\ell)}_{j}(\sigma)\,
\frac{d \sigma}{\sigma}
\end{equation}
stands for the $j$--th~order $\ell$--loop ``timelike'' effective couplant,
that constitutes the proper continuation of the $j$--th power of
$\ell$--loop QCD couplant~$\bigl[a\ind{(\ell)}{s}(Q^2)\bigr]^{j}$ into the
timelike domain.

As~discussed in Sect.~\ref{Sect:Repem1L}, at the one--loop level the
first--order function~(\ref{ATLDef}) acquires a quite simple form,
namely~(\ref{RCTL1L})
\begin{equation}
\label{ATL1L1pExpl}
\ATL{(1)}{1}(s) = a\inds{(1)}{TL}(s) =
\frac{1}{2} - \frac{1}{\pi}\arctan\!\left(\frac{\ln w}{\pi}\right)\!\!,
\quad w = \frac{s}{\Lambda^2}.
\end{equation}
Despite the fact that the partial spectral
functions~$\bar\rho^{(\ell)}_{j}(\sigma)$~(\ref{RhoPartHO}) become rather
cumbrous at the higher loop levels, it appears that the integration in
Eq.~(\ref{ATLDef}) can be performed explicitly for~$\ell>1$, too. For
example, the two--loop first--order function~(\ref{ATLDef})
reads~\cite{Rad82}
\begin{equation}
\label{ATL2L1pExpl}
\ATL{(2)}{1}(s) = a\inds{(1)}{TL}(s) - \frac{B_{1}}{\ln^{2}w+\pi^2}\!
\left[ W(s) - a\inds{(1)}{TL}(s) \ln w + 1 \right]\!,
\end{equation}
whereas the second--order one can be represented as
\begin{align}
\label{ATL2L2pExpl}
\ATL{(2)}{2}(s) & = \frac{1}{\ln^{2}w+\pi^2} +
\frac{B_{1}}{\left(\ln^{2}w+\pi^2\right)^2} \times
\nonumber \\ & \times\!
\biggl\{\!
a\inds{(1)}{TL}(s)\Bigl(\ln^{2}w-\pi^2\Bigr) - \ln w\,\Bigl[2W(s)+1\Bigr]\!\!\biggr\} +
\nonumber \\
& + \frac{B_{1}^{2}}{\left(\ln^{2}w+\pi^2\right)^{\!\!3}}
\Biggl\{\!\!\left(\ln^{2}w - \frac{\pi^2}{3}\right)\!\! \times
\nonumber \\ & \times\!\!
\left[\!\!\left(\!W(s)+\frac{1}{3}\right)^{\!\!2}\!\! +
\frac{1}{9} -
\pi^2\left[a\inds{(1)}{TL}(s)\right]^{2}\right] -
\nonumber \\[1.25mm]
& -\frac{2}{3}a\inds{(1)}{TL}(s)\ln w \Bigl(\ln^{2}w - 3\pi^2\Bigr)
\!\!\left(\!W(s)+\frac{1}{3}\right)\!\!\Biggr\}.
\end{align}
In these equations~$a\inds{(1)}{TL}(s)$ is given by
Eq.~(\ref{ATL1L1pExpl}), $B_j=\beta_j/\beta^{j+1}_{0}$, and
\begin{equation}
\label{WsDef}
W(s) = \ln\sqrt{\ln^{2}w + \pi^2},
\qquad
w = \frac{s}{\Lambda^2}.
\end{equation}

\begin{figure}[t]
\centering
\includegraphics[width=75mm,clip]{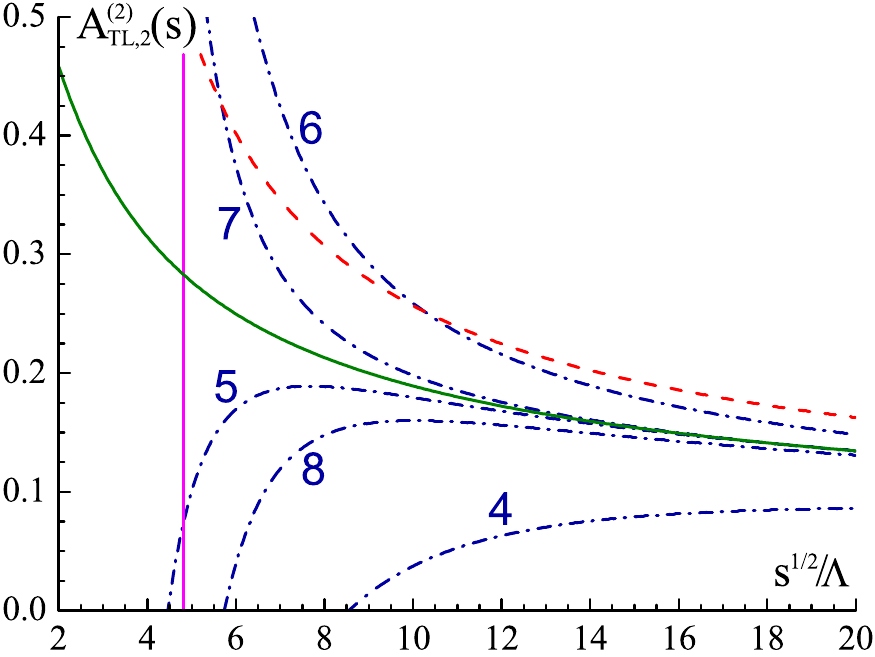}
\caption{Two--loop timelike effective expansion function of the second
order~$\ATLs{(2)}{2}(s)$ [Eq.~(\ref{ATL2L2pExpl}), solid curve] and the
approximations corresponding to various orders of its
re--expansion~(\ref{ATL2L2pUV}). The result of ``naive''
continuation~(\ref{RNaive}) [the first term on the right--hand side of
Eq.~(\ref{ATL2L2pUV})] is shown by dashed curve. Numerical label next to a
dot--dashed curve indicates the highest absolute value of power of~$\ln w$
retained on the right--hand side of Eq.~(\ref{ATL2L2pUV}). The plotted
functions are scaled by the factor of~$10$. Vertical solid line marks the
boundary of convergence range of the re--expansion~(\ref{ATL2L2pUV}) at
$\sqrt{s}/\Lambda = \exp(\pi/2) \simeq 4.81$.}
\label{Plot:ATL2L2pUV}
\end{figure}

In turn, at the three--loop level the first--order function~(\ref{ATLDef})
takes the form
\begin{align}
\label{ATL3L1pExpl}
\ATL{(3)}{1}(s) & = \ATL{(2)}{1}(s) - \frac{B_{1}^2}{\left(\ln^{2}w+\pi^2\right)^{\!\!2}}
\Bigl[T_{1}(s)T_{2}(s) + \ln w \Bigr] +
\nonumber \\ & +
\frac{B_{2}\ln w}{\left(\ln^{2}w+\pi^2\right)^{\!\!2}},
\end{align}
where~$\ATL{(2)}{1}(s)$ is specified in Eq.~(\ref{ATL2L1pExpl}) and
\begin{align}
\label{T1Def}
T_{1}(s) & = a\inds{(1)}{TL}(s)\,\ln w - W(s), \\[1.5mm]
\label{T2Def}
T_{2}(s) & = \pi^2 a\inds{(1)}{TL}(s) + W(s)\,\ln w.
\end{align}
As for the four--loop first--order function~(\ref{ATLDef}), it can be
represented~as
\begin{align}
\label{ATL4L1pExpl}
\ATL{(4)}{1}(s) & = \ATL{(3)}{1}(s) +
\frac{1}{\left(\ln^{2}w+\pi^2\right)^{\!\!3}}\frac{B_{3}}{2}\!
\left(\ln^{2}w-\frac{\pi^2}{3}\right) +
\nonumber \\ & \hspace{-7mm}  +
\frac{B_{1}B_{2}}{\left(\ln^{2}w+\pi^2\right)^{\!\!3}}\!
\biggl\{\frac{\pi^2}{3} - 3T_{2}(s)\ln w +\pi^2 W(s)+
\nonumber \\ & \hspace{-7mm} + \ln^{2}w\Bigl[a\inds{(1)}{TL}(s)\ln w-1\!\Bigr]\!\!\biggr\} -
\frac{1}{\left(\ln^{2}w+\pi^2\right)^{\!\!3}}\frac{B_{1}^3}{2}\times
\nonumber \\ & \hspace{-7mm} \times\!\!
\Biggl\{\!2T_{2}(s)\ln w \Bigl[W(s)-3\Bigr]
-T_{2}^2(s)\!\Bigl[2T_{1}(s)+1\Bigr]\! +
\nonumber \\ & \hspace{-7mm} +
\pi^2T_{1}^2(s)\biggl[\frac{2}{3}T_{1}(s)+3\biggr] +
\ln^{2}w\Bigl[2\,a\inds{(1)}{TL}(s)\ln w-1\Bigr] +
\nonumber \\ & \hspace{-7mm} +
2W^{2}(s)\ln^{2}w\Bigl[T_{1}(s)+W(s)-2\Bigr] + \frac{\pi^2}{3} +
\nonumber \\ & \hspace{-7mm} +
2W(s)\Bigl[1-W(s)\Bigr]\!\!\Bigl[\pi^2+a\inds{(1)}{TL}(s)\ln^{3}w\Bigr]\!\!\Biggr\},
\end{align}
where the functions~$\ATL{(3)}{1}(s)$, $W(s)$, $T_{1}(s)$, and~$T_{2}(s)$
are given by Eqs.~(\ref{ATL3L1pExpl}), (\ref{WsDef}), (\ref{T1Def}),
and~(\ref{T2Def}), respectively. It is worthwhile to note also that the
functions~$\ATL{(\ell)}{j}(s)$~(\ref{ATLDef}) entering Eq.~(\ref{RProp2})
can be computed numerically by making use of the routines included in the
freely available program packages~\cite{QCDDPT1} [which, being based on a
less universal method of calculation of the relevant spectral
function~(\ref{RhoDef}) than that of Eq.~(\ref{RhoPertHO}), is applicable
at first four loop levels only] and~\cite{QCDDPT2}.

\begin{figure}[t]
\centering
\includegraphics[width=75mm,clip]{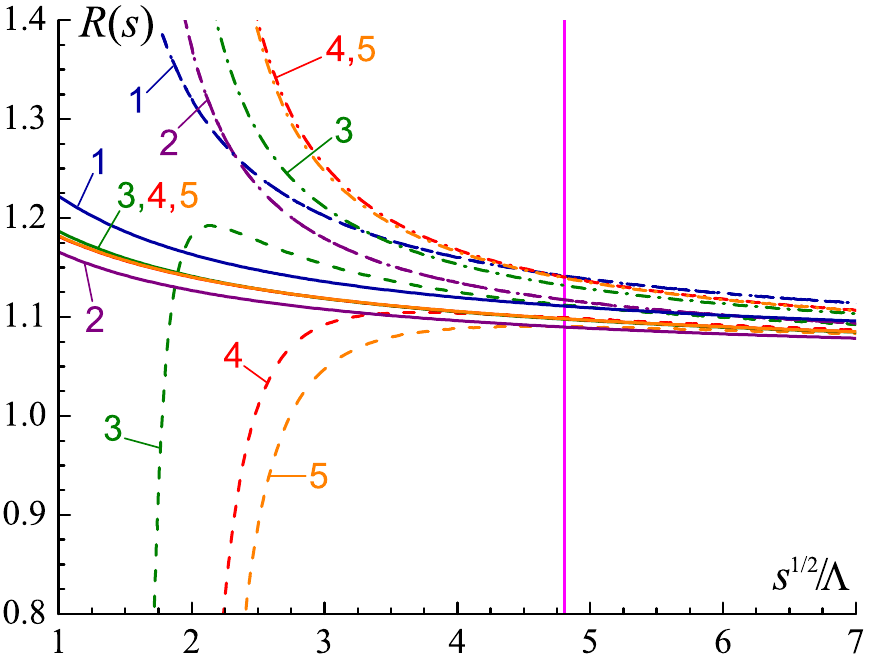}
\caption{The function~$R^{(\ell)}(s)$ [Eq.~(\ref{RProp}), solid curves],
its naive form [Eq.~(\ref{RNaive}), dot--dashed curves], and the
re--expanded approximation [Eq.~(\ref{RAppr}), dashed curves] at various
loop levels ($1 \le \ell \le 5$). Vertical solid line marks the boundary
of convergence range of the re--expansion~(\ref{RAppr}) at
$\sqrt{s}/\Lambda = \exp(\pi/2) \simeq 4.81$. Numerical labels specify the
loop level.}
\label{Plot:R5L}
\end{figure}

As discussed earlier, the~$\pi^2$--terms play a valuable role in the
studies of the strong interaction processes in the timelike domain, and
their ignorance (complete or partial) may yield misleading results.
In~particular, this issue is illustrated by Fig.~\ref{Plot:ATL2L2pUV},
which displays the two--loop timelike effective expansion function of the
second order~$\ATL{(2)}{2}(s)$ [Eq.~(\ref{ATL2L2pExpl}), solid curve] and
the approximations (dashed and dot--dashed curves) corresponding to
various orders of its re--expansion for~$\sqrt{s}/\Lambda > \exp(\pi/2)
\simeq 4.81$:
\begin{align}
\label{ATL2L2pUV}
\ATL{(2)}{2}(s) & \simeq \left[a\ind{(2)}{s}(|s|)\right]^{2}
\!\!- \frac{\pi^2}{\ln^{4}w}
+ \frac{\pi^2}{\ln^{5}w}\,B_{1}\!\!\left(\!4\ln\ln w-\frac{7}{3}\right)\! -
\nonumber \\
& - \frac{\pi^2}{\ln^{6}w}\!\left[\frac{B_{1}^{2}}{3}\Bigl(\!
10\ln^{2}\ln w - 9\ln\ln w + 1\!\Bigr)\! - \pi^2\right]\! -
\nonumber \\
& - \frac{\pi^4}{\ln^{7}w}\frac{3B_{1}}{10}\Bigl(20\ln\ln w - 19\Bigr)\! +
\co\!\left(\frac{1}{\ln^{8}w}\right)\!\!.
\end{align}
Specifically, the dashed curve shows the result of naive continuation of
the respective term of the Adler function perturbative expansion into the
timelike domain~(\ref{RNaive}) [the first term on the right--hand side of
Eq.~(\ref{ATL2L2pUV})], whereas the dot--dashed curves additionally
include\footnote{Note that all the terms of the
re--expansion~(\ref{ATL2L2pUV}) (except for the naive one) appear to be
discarded in the two--loop approximate
expression~$R^{(2)}_{\text{appr}}(s)$~(\ref{RAppr}).} the lowest--order
$\pi^2$--terms [numerical labels indicate the highest absolute value of
power of~$\ln w$ retained in Eq.~(\ref{ATL2L2pUV})].
Figure~\ref{Plot:ATL2L2pUV} implies that the
re--expansion~(\ref{ATL2L2pUV}) converges rather slowly at low and
moderate energies. Furthermore, even at relatively high energies
$\ATL{(2)}{2}(s)$~(\ref{ATL2L2pExpl}) considerably differs
from~$\bigl[a\ind{(2)}{s}(|s|)\bigr]^{2}$, the latter being the only part
of the function~(\ref{ATL2L2pExpl}), which is retained in the two--loop
approximation of the $R$--ratio~(\ref{RAppr}). For~example, as one can
infer from Fig.~\ref{Plot:ATL2L2pUV}, for~$\sqrt{s}/\Lambda=20$ the
function $\bigl[a\ind{(2)}{s}(|s|)\bigr]^{2}$ exceeds~$\ATL{(2)}{2}(s)$ by
about~$21\,\%$, and to securely achieve~$10\,\%$ accuracy in the
re--expansion~(\ref{ATL2L2pUV}) the inclusion of the~$\pi^2$--terms up to
the order of~$\ln^{-7}w$ is required.

The issue of the higher--loop stability of the $R$--ratio of
electron--positron annihilation into hadrons is also elucidated by
Fig.~\ref{Plot:R5L}. In particular, this Figure displays the proper
expression~$R^{(\ell)}(s)$ [Eq.~(\ref{RProp}), solid curves], its naive
form~$R^{(\ell)}_{\text{naive}}(s)$ [Eq.~(\ref{RNaive}), dot--dashed
curves], and the truncated re--expanded
approximation~$R^{(\ell)}_{\text{appr}}(s)$ [Eq.~(\ref{RAppr}), dashed
curves] at various loop levels ($1 \le \ell \le 5$). Figure~\ref{Plot:R5L}
makes it evident that the loop convergence of the commonly employed
approximation~$R^{(\ell)}_{\text{appr}}(s)$~(\ref{RAppr}) is worse than
that of both the proper expression~$R^{(\ell)}(s)$~(\ref{RProp}) and the
naive one~$R^{(\ell)}_{\text{naive}}(s)$~(\ref{RNaive}). As discussed
earlier, this is primarily caused by the fact that the convergence range
of the re--expanded approximation~(\ref{RAppr}) is strictly limited
to~$\sqrt{s}/\Lambda > \exp(\pi/2) \simeq 4.81$, that, in turn, results in
rather large values of the higher--order
coefficients~$\delta_j$~(\ref{rjDef}) embodying the contributions of the
corresponding~$\pi^2$--terms~(\ref{Pi2TermsGen}). In particular, as one
can infer from Fig.~\ref{Plot:R5L}, beyond the two--loop level (i.e., for
$\ell \ge 3$) the curves corresponding to $R^{(\ell)}(s)$~(\ref{RProp})
are nearly indistinguishable from each other, whereas the curves
corresponding to $R^{(\ell)}_{\text{appr}}(s)$~(\ref{RAppr}) start to
swerve quite above the boundary of its convergence range. For example, at
$\sqrt{s}/\Lambda = 2\exp(\pi/2) \simeq 9.62$ the relative difference
between the $\ell$--loop and $(\ell+1)$--loop strong corrections to the
proper expression $R^{(\ell)}(s)$~(\ref{RProp}) is~$0.4\,\%$ for~$\ell=3$
and~$0.003\,\%$ for~$\ell=4$, whereas for the case of the commonly
employed approximation~$R^{(\ell)}_{\text{appr}}(s)$~(\ref{RAppr}) these
values increase up to~$4.4\,\%$ for~$\ell=3$ and~$2.5\,\%$ for~$\ell=4$.

\begin{figure}[t]
\centering
\includegraphics[width=75mm,clip]{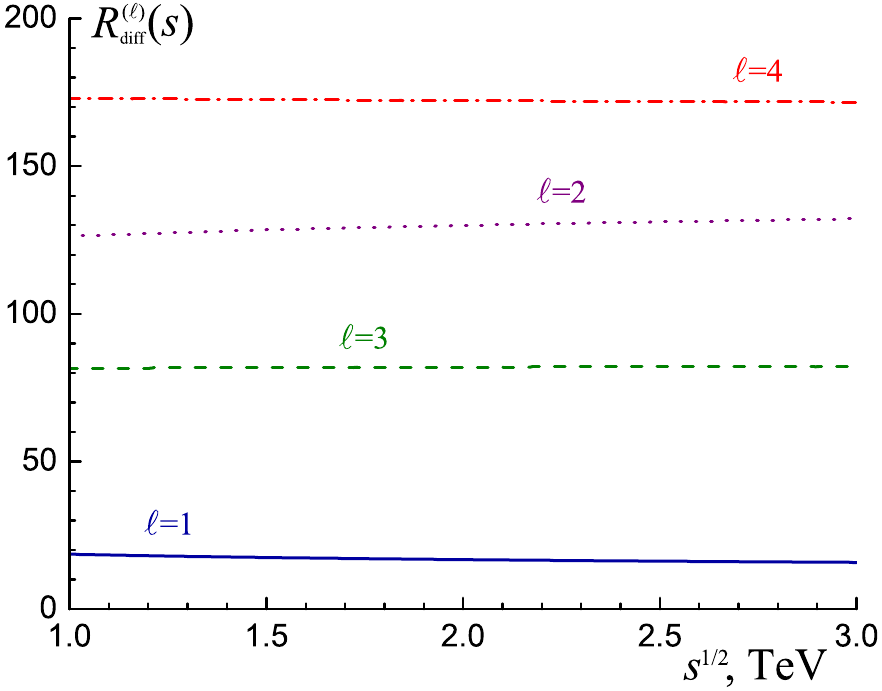}
\caption{The function~$R\indt{(\ell)}{\text{diff}}(s)$~(\ref{RDiff}) in
the energy range planned for the future CLIC experiment~\cite{CLIC} at
various loop levels.}
\label{Plot:R5LDiff}
\end{figure}

To expound the accuracy of approximation of the $R$--ratio~(\ref{RProp})
by its truncated
re--expansion~$R^{(\ell)}_{\text{appr}}(s)$~(\ref{RAppr}), it is
worthwhile to mention also the following. At a moderate energy scale of
the mass of $\tau$~lepton $\sqrt{s}=M_{\tau}$ the relative difference
between the $\ell$--loop strong corrections $r^{(\ell)}(s)$~(\ref{RProp})
and $r^{(\ell)}_{\text{appr}}(s)$~(\ref{RAppr}) turns out to be as high
as~$26\,\%$, $28\,\%$, $14\,\%$, $2\,\%$, and~$7\,\%$ at the one--, two--,
three--, four--, and five--loop levels, respectively. Moreover, it appears
that even at high energies the ignorance of the higher--order
$\pi^2$--terms in the truncated re--expanded approximation~(\ref{RAppr})
may produce a considerable effect. In particular, this issue is
illustrated by Fig.~\ref{Plot:R5LDiff}, which displays the quantity
\begin{equation}
\label{RDiff}
R^{(\ell)}_{\text{diff}}(s) =
\left|\frac{R^{(\ell)}_{\text{appr}}(s)-R^{(\ell)}(s)}
{R^{(\ell)}_{\text{appr}}(s)-R^{(\ell+1)}_{\text{appr}}\!(s)}\right|\!\times\! 100\,\%
\end{equation}
at various loop levels. Specifically, as one can infer from
Fig.~\ref{Plot:R5LDiff}, in the energy range planned for the future CLIC
experiment~\cite{CLIC} the effect of inclusion of the $\pi^2$--terms
discarded in the approximate expression
$R\ind{(\ell)}{appr}(s)$~(\ref{RAppr}) is either comparable to or
prevailing over the effect of inclusion of the next--order perturbative
correction.

\section{Conclusions}
\label{Sect:Concl}

The strong corrections to the $R$--ratio of electron--positron
annihilation into hadrons are studied at the higher--loop levels.
In~particular, the derivation of a general form of the commonly employed
approximate expression for the $R$--ratio (which constitutes its truncated
re--expansion at high energies) is delineated, the appearance of the
pertinent $\pi^2$--terms is expounded, and their basic features are
examined. It~is demonstrated that the validity range of such approximation
is strictly limited to $\sqrt{s}/\Lambda > \exp(\pi/2) \simeq 4.81$ and
that it converges rather slowly when the energy scale approaches this
value. The~spectral function required for the proper calculation of the
$R$--ratio is explicitly derived and its properties at the higher--loop
levels are studied. The~developed method of calculation of the spectral
function enables one to obtain the explicit expression for the latter at
an arbitrary loop level. By making use of the derived spectral function
the proper expression for the $R$--ratio is calculated up to the
five--loop level and its properties are examined. In~particular, it is
shown that the loop convergence of the proper expression for the
$R$--ratio is better than that of its commonly employed approximation and
that the omitted higher--order $\pi^2$--terms in the latter may produce a
considerable effect.

\begin{acknowledgements}
The author is grateful to A.B.~Arbuzov for the stimulating discussions and
useful comments.
\end{acknowledgements}

\end{document}